\documentclass[aps,prb,twocolumn]{revtex4-2}
\usepackage{graphicx}
\usepackage{hyperref}
\usepackage{xcolor}
\usepackage{amsmath}
\usepackage{amsfonts}
\usepackage{amssymb}
\usepackage{varioref}
\usepackage{float}
\usepackage{dcolumn}   
\usepackage{caption}
\usepackage{subcaption}
\usepackage{epstopdf}
\usepackage{braket}

\begin{document}
	\textheight=23.8cm

\title{ Driven one-dimensional noisy Kitaev chain}

\author{Manvendra Singh, Santanu Dhara, and Suhas Gangadharaiah}
\affiliation {Department of Physics, Indian Institute of Science Education and Research, Bhopal, India}

\date{\today}

\begin{abstract}
We study  one dimensional Kitaev chain driven by the anisotropy parameter, $J_{-}= t/\tau_Q$, in the presence of weak Gaussian noise in the parameter $J_{+}$. The system is prepared in the ground state of the Hamiltonian at the initial time $t\rightarrow -\infty$. 
The defect density and the residual energy at the end of the drive protocol reveals anti-Kibble Zurek (AKZ) behavior which deviates from the earlier reported linear in quench time dependence for the defect density. The entropy density at the end of the protocol is found to exhibit the signature of the AKZ behavior. 
In the context of the Kitaev chain, the two-point spin correlators are short ranged. The hidden topological order across the quantum phase transition point is probed via the non-local string  order parameters. The  two-point Majorana correlator and the hidden string correlator calculated in the final decohered state at the end of the noisy drive protocol is consistent  with the AKZ picture. We have analyzed the kink statistics in the dual spin space, the first three cumulants at the end of the noisy drive also exhibit the AKZ scaling behavior for slower sweeps similar to the defect density. The kink distribution function is well approximated with the normal distribution with non-universal noise dependent mean and variance.
\end{abstract}

\maketitle

\section{Introduction}
In the context of the drive through a second order phase transition, the Kibble-Zurek mechanism (KZM) predicts the production of topological defects across the phase transition point  where the defect density, $n$, scales with the quench time $\tau_Q$  as  $n\sim \tau^{-\beta}_Q$, with the  KZM exponent $\beta=d\nu/(1+\nu z)$ expressed  in terms of the universal critical exponents $\nu$ and $z$ which are the  correlation length and  the dynamical exponent, respectively, while $d$ is the dimension of the system~\cite{Kibble_1976, KIBBLE1980183, Zurek1985, Kibble1985, Zurek1994, ZUREK1996177}. The applicability of the KZM has been extended to describe  non-equilibrium quantum  dynamics in the context of quantum phase transition (QPT) in various scenarios ~\cite{PhysRevLett.95.035701, PhysRevLett.95.105701, PhysRevLett.95.245701, PhysRevB.72.161201, RevModPhys.83.863, doi:10.1142/S0217751X1430018X,Lamporesi2013, Cui2016, Keesling2019}. Even for infinitesimally slow drive the creation of the defects  across the quantum critical point occurs due to the vanishing energy gap and diverging relaxation time at the quantum critical point (QCP) i.e., the system is unable to follow the instantaneous ground state in the drive through the QCP resulting in the creation of excitations. 

Recent studies~\cite{PhysRevX.2.041022, YUKALOV20151366, PhysRevLett.117.080402, PhysRevB.95.224303,Weinberg2020,PhysRevX.7.041014, PhysRevLett.124.230602, PhysRevB.101.104307, PhysRevA.103.012608, PhysRevB.104.064313,King2022} have established that the drive protocols through the QCP that have noisy component to it  are more susceptible towards defect generation and increased residual energy specifically in the slow sweep regime. In particular, it was shown numerically for the transverse field quantum Ising model (TFQIM)~\cite{PhysRevLett.117.080402} and the XY-spin chain model~\cite{PhysRevB.95.224303}, that for certain protocols  the defect density can be approximated as,  $n\approx c \tau_Q^{-\beta}+d \eta^2_0 \tau_Q $ for the weak Gaussian noise case, with $c$ and $d$ being system dependent  parameters and $\eta_0$ the noise strength. The result was further corraborated by the analytical studies on XY-spin chain with transverse drive protocol~\cite{PhysRevB.104.064313}.  The  linear in $\tau_Q$ term  owes its origin to the presence of noise
and is the Anti-Kibble-Zurek (AKZ) term which is responsible for the increased defect generation in the slow-sweep regime.  In the present work, we study the driven one dimensional Kitaev chain across a QCP. The protocol considered here is qualitatively different from those considered in ~\cite{PhysRevLett.117.080402, PhysRevB.104.064313} and results in significantly different behavior for the AKZ term. For the noisy Kitaev chain driven by the Hamiltonian parameter $J_{-}$,  we find that the noise dependent defect density has linear (in quench time) term multiplied by a log-term in the weak noise limit a result different from all the earlier studies~\cite{PhysRevLett.117.080402,PhysRevB.95.224303,PhysRevB.104.064313}. As a consequence, the optimal quench time has a completely different dependence on the noise strength from the previously reported $\eta^{-4/3}$ power.
However, the residual energy exhibits conventional behavior with the noise dependent term being linear in quench time.

The consequences of the (noiseless) drive through the QCP and the KZM picture of defect formation is also reflected in the von Neumann entropy density~\cite{PhysRevA.73.043614}.
In the KZM scenario, slower sweep results in lower entropy density. In  both the fast and the slow sweep regimes of the noiseless drive, the entropy density is strongly suppressed while the entropy density is maximized  at an intermediate sweep speed.
However, for the noisy drive case, there is a characteristic quench time scale related to the optimal quench time  beyond which the entropy density increases.  Interestingly, the entropy density also signifies the crossover behavior of the correlation functions~\cite{PhysRevA.73.043614, PhysRevA.75.052321, PhysRevB.78.045101}.

For the Kitaev chain,  the two point spin-spin correlators are short ranged (i.e., the spin correlators vanishes for more than  one lattice constant separation which signifies that both phases across the QCP are disordered)~\cite{PhysRevLett.98.247201, PhysRevLett.98.087204,  Chen_2008}. 
 Unlike the prediction from Landau's theory of the second order continuous phase transition,  a QPT can also occur between the two disordered phases without any symmetry breaking related to the change of topological order. Many low dimensional quantum systems 
exhibit such QPT of topological nature without any symmetry breaking associated with the local order parameter (LOP). Examples include topological and Mott insulators,  spin liquids,  frustrated magnetic systems, etc.~\cite{PhysRevLett.98.087204, PhysRevLett.98.247201, Chen_2008, PhysRevLett.109.236404, Ryu_2010, Chitov_2017, PhysRevB.100.104428,  Bahovadinov_2019, dutta_aeppli_chakrabarti_divakaran_rosenbaum_sen_2015, PhysRevD.17.2637}. Therefore, the conventional Landau theory based on the LOP is not directly applicable to characterize such QPT.
In the quantum phase transitions of the topological type, the hidden topological order across the QCP  can  in principle  be probed via various non-local string order parameters  (SOP) by extending the Landau theory to accommodate these non-local string operators and their correlation functions as suggested in various works~\cite{Chitov_2017, PhysRevB.100.104428, Bahovadinov_2019}. These non-local SOP show either long range order or short range order across the QCP thus successfully describing hidden order or disorder in different phases. Motivated from these studies, we calculate  the two-point Majorana correlator and a hidden string operator to probe hidden order or disorder for the driven one dimensional noisy Kitaev chain. In the noiseless drive case, both the two-point Majorana correlator and a hidden string correlator show long range order for slower drives which is consistent with KZM predictions~\cite{PhysRevLett.100.077204, PhysRevB.78.045101, PhysRevLett.98.087204}. Furthermore, the two-point Majorana correlator and  the non-local hidden SOP exhibit the signature of the AKZ behavior, i.e., correlation length-scale related to the ordered domain quantified by these non-local correlators decreases due to the presence of noise.

 Next, we considered the kink/defect distribution in the dual spin space at the end of the noisy drive protocol. In particular, we calculate the first three cumulants (mean, variance and third central moment), all of which exhibit the universal KZ scaling behavior ($\propto 1/\sqrt{\tau_Q}$) with the quench time in the slower drive regime for the noiseless case as discussed in Ref.~\cite{Campo_2018}. For the noisy drive cases we find that all the three cumulants  exhibit the KZ as well as the noise dependent AKZ scaling behavior similar to what we find for the defect density. Similar to the noiseless case, the kink distribution for the noisy drives in the slow sweep regime approaches the normal distribution $\mathcal{N}(\kappa_1,\kappa_2)$ having non-universal noise dependent mean ($\kappa_1=\langle n \rangle$) and variance ($\kappa_2 = \langle n^2 \rangle - \langle n \rangle^2$), however unlike the noiseless drive case the variance and other higher cumulants no longer remain proportional to the kink density.

The outline of this article is as follows. In  Sec.~\ref{MHam},  we introduce the model Hamiltonian for the one dimensional  Kitaev model and study the dynamics of the system for the case of the noisy drive through the QCP in the thermodynamic and in the long time limit.
We solve the dynamics of the system by mapping the time dependent Hamiltonian in momentum space to the set of independent noisy Landau-Zener (LZ) problems.  In Sec.~\ref{DDRE}, we calculate the defect density and the residual energy which show AKZ behavior for the noisy drive protocol. In Sec.~\ref{EDensity} we calculate the entropy density for the final decohered state due to noisy drive through the QCP. In Sec.~\ref{StrCorr}, we study the non-local string operator correlators, in particular we calculate the two-point Majorana correlator and a hidden SOP  (at finite separation) to characterize the hidden (topologically) ordered or disordered phase of the Kitaev chain at the end of the noisy drive. These non-local correlators also exhibit the signature of AKZ picture of defect generation in the noisy drive protocol. In Sec.~\ref{KStat} we show that the lower order cumulants corresponding to the kink distribution function exhibit AKZ behavior similar to that in the defect density.  We conclude our results in the Sec.~\ref{Disc}.
\section{Model Hamiltonian}\label{MHam}
The one-dimensional Kitaev model Hamiltonian is given by the Hamiltonian ~\cite{PhysRevB.78.045101, PhysRevB.79.224408,PhysRevB.81.054306,KITAEV20062},
\begin{equation}
    H=\sum^{N/2}_{j=1} \left(J_1\,\sigma^x_{2j} \sigma^x_{2j+1}+J_2\,\sigma^y_{2j-1} \sigma^y_{2j}\right),
    \label{eq:H_spin}
\end{equation}
where $\sigma^{x,y}$ are the Pauli matrices,  and the parameters $J_1$ and $J_2$ are the bond dependent spin couplings. Performing the Jordan-Wigner transformation with,
\begin{equation}
    a_j= (\Pi^{2j-1}_{i=-\infty}\sigma^z_i)\sigma^y_{2 j}\,\,,\,\,b_j= (\Pi^{2j}_{i=-\infty}\sigma^z_i)\sigma^x_{2 j+1},
\label{eq:JW}
\end{equation}
where $a_j$ and $b_j$ are the two independent Majorana fermions ( i.e., $a^{\dagger}_j=a_j$, $b^{\dagger}_j=b_j$)
at site $j$  satisfying $\{a_i,a_j\}=2\delta_{ij}$ , $\{b_i,b_j\}=2\delta_{ij}$ and $\{a_i,b_j\}=0$. 
The Hamiltonian given by Eq.~\ref{eq:H_spin}  expressed in terms of the Fourier transformed variables acquires the following form~\cite{dutta_aeppli_chakrabarti_divakaran_rosenbaum_sen_2015},
\begin{equation}
    H=\sum^{\pi}_{k=0} \psi^{\dagger}_k \mathcal{ H}_k \psi_k
\end{equation}
with,
\begin{equation}
   \mathcal{ H}_k=2 i \begin{bmatrix}
  0&  -J_1 -J_2 e^{-i k} \\
J_1 +J_2 e^{i k} & 0 
\end{bmatrix},\\
\label{eq:H_matrix1}
\end{equation}
where $\psi_k=(a_k, b_k)^T$, note $a_k$ and $b_k$ are Fourier components which satisfy the standard anti-commutation relations.
We consider the quenching scheme wherein,  $J_{-} =J_1-J_2 $,  is varied linearly with time as:  $J_{-}(t) =t/\tau_Q$, 
with $t$ running from $-\infty$ to $\infty$ and $J_{+}=J_1+ J_2$.
We next perform unitary transformation so as to recast the Hamiltonian $\mathcal{H}_k$  in Eq.~[\ref{eq:H_matrix1}] into the Landau-Zener type Hamiltonian of the following form
\begin{equation}
   \tilde{ \mathcal{H}}_k=2\, \begin{bmatrix}
  J_{-}(t)\sin(k/2)&  J_{+}\cos(k/2)  \\
J_{+}\cos(k/2) & -J_{-}(t)\sin(k/2) 
\end{bmatrix}\\, 
\label{eq:H_matrix2}
\end{equation}
where the $k$ runs from $0$ to $\pi$. The energy gap, $ 4 \sqrt{J^2_{-}\sin^2{\frac{k}{2}}+J^2_{+}\cos^2{\frac{k}{2}}}$, vanishes at the critical point $J_{-}=0$ for the critical mode $k=\pi$, and corresponds to the  quantum phase transition of topological nature with the critical exponents, $\nu=z=1$~\cite{PhysRevLett.98.087204, KITAEV20062}. Note that although the Kitaev spin chain can be mapped to TFQIM in the dual spin space,  however the quenching scheme considered here which involves $J_{-}$ leads to the fundamentally different fermionized Hamiltonian as compared to that of the Hamiltonian of the TFQIM~\cite{PhysRevLett.117.080402} or the TFQIM  limit of the XY spin chain case considered for  the transverse protocol~\cite{PhysRevB.104.064313}.

Apart from the linear drive term $J_{-}(t)=v t=t/\tau_Q$, we consider  Gaussian noise dependence in $J_{+}$  i.e.,  $J_{+}(t)=J_{+}+\eta(t)$, the noisy LZ type Hamiltonian can thus be written as, 
\begin{equation}
    \mathcal{H}^{\eta}_k(t)=  \tilde{ \mathcal{H}}_k+ [2 \eta(t) \cos{\frac{k}{2}}] \tau_1,
    \label{eq:H_matrix_noisy}
\end{equation}
where $\tau_1$ is the $x$-Pauli matrix. The  noise term $\eta(t)$ is Gaussian correlated  and characterized by  zero mean 
and $\overline{\eta(t) \eta(t_1)} =\eta_0^2 e^{-\Gamma |t-t_1|}$, where $\eta_0$ is the noise strength and $\Gamma$ is the inverse time scale  corresponding to the noise. We would like to point out that the  noise term does not change the original translational invariance of the problem. Therefore, the  k-modes remain  decoupled even in the presence of noise.   Moreover, since  the initial state that we consider has a tensor product structure,  the k-modes remain decoupled (since there are neither interactions nor impurities present), a statement which holds even under noise averaging. A second point we will briefly discuss here is the difference between the drive protocol considered for the noisy Kitaev chain and the quench protocol considered in $XY$-spin chain~\cite{PhysRevB.104.064313} while taking the $\gamma=1$ limit and the noisy TFQIM considered in~\cite{PhysRevLett.117.080402}.
As discussed in Sec.~\ref{StrCorr},
the duality transformation $\sigma^x_j =\tau^x_{j-1} \tau^x_j$ and $\sigma^y_j=\Pi^{N}_{l=j} \tau^{y}_l$, 
maps the Kitaev Hamiltonian   to the TFQIM, i.e., $H_d= \sum^{N/2}_{j=1} (J_1 \tau^{x}_{2j-1} \tau^x_{2j+1}+J_2 \tau^y_{2j-1})$. Since we consider the drive by $J_{-}$ while the noise is present in the $J_{+}$ term, therefore both the spin terms include the driving and the noise term. While in our model the drive and the noise term in the $J_1$ term come with the same relative sign, in the $J_2$ term they come with the opposite signs. On the other hand, in the TFQIM considered in~\cite{PhysRevLett.117.080402} both the 
$J_1$ and $J_2$ have the noise and the drive terms with the same relative sign, in other words the driving term has noise in it. For the TFQIM obtained by the $\gamma=1$ limit of the $XY$-spin chain~\cite{PhysRevB.104.064313}, as opposed to the Kitaev model, the noise term is present only in the spin-spin term while the drive parameter is on the transverse term. 

\begin{figure}
    \centering
    \includegraphics[width=\columnwidth]{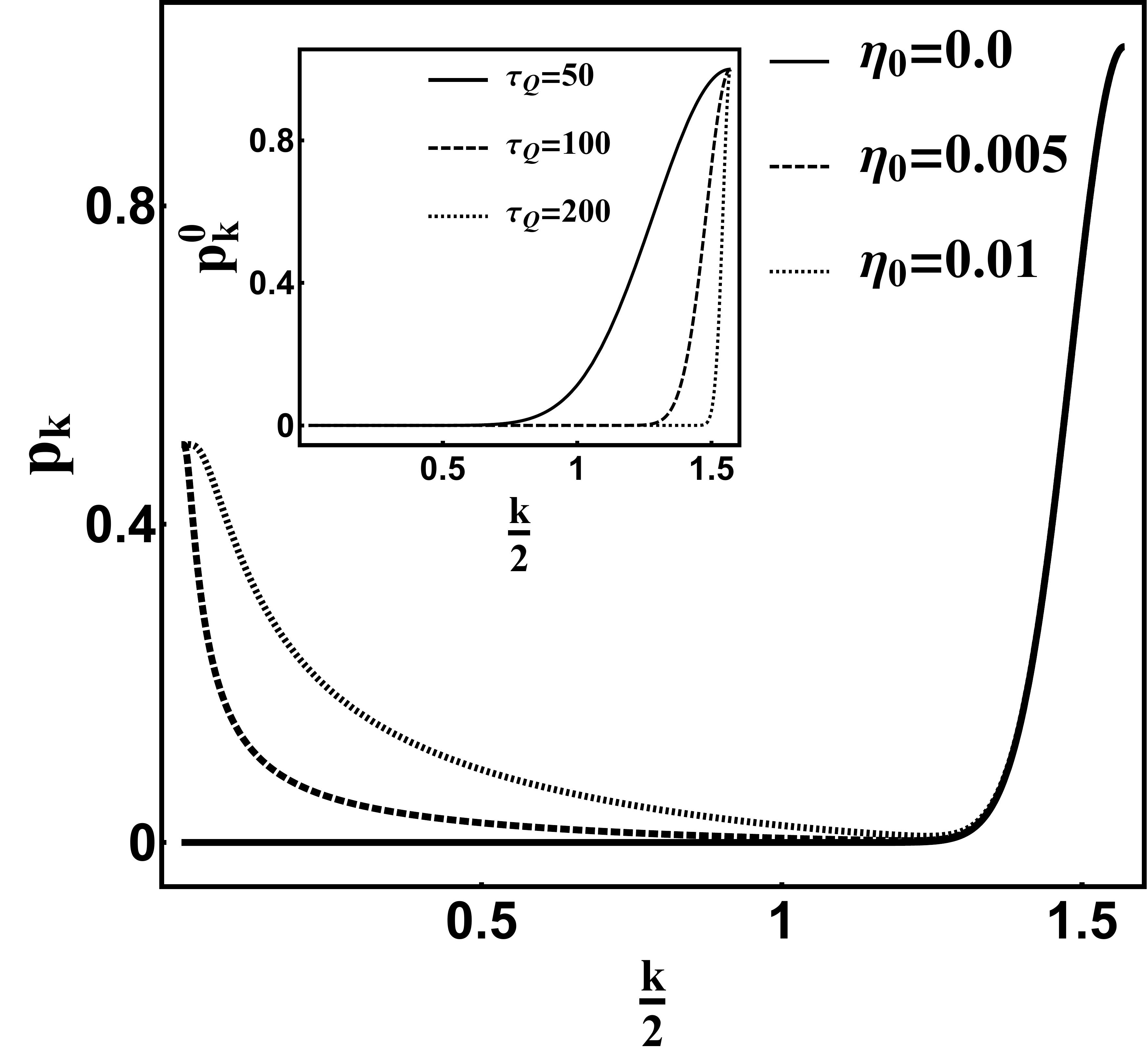}
    \caption{In the inset we have plotted the excitation probability $p^{0}_{k}$ as a function of $k$ for the noiseless drive scenario.  For the slower drive, i.e., large $\tau_Q$ case, the excitation probability is non-zero around the critical point $k=\pi$ only. This  suggests that the non-adiabatic contribution to the defect density comes around the critical point only in the slower sweep regime. In the main figure we have plotted the excitation probability $p_{k}$ vs. $k$, for $\tau_Q=100$ (which is the slow drive regime), in the presence of noisy drive. 
    The increase of the noise strength, $\eta_{0}$ opens up a new non-adiabatic region around the $k=0$ point.}
    \label{fig:pk0}
\end{figure}

A general time evolved state $\ket{\psi^{\eta}_k(t)}$ can be written as, $\ket{\psi^{\eta}_k(t)} =c^{\eta}_{1k}(t)\ket{\Bar{e}_{1k}}+c^{\eta}_{2k}(t)\ket{\Bar{e}_{2k}}$, for a single realisation of the Gaussian noise,
where $c^\eta_{i k}(t)$ are time dependent probability amplitudes for the (diabatic) states $\ket{\Bar{e}_{i k}}$ (which are the eigenstates of Hamiltonian, Eq.~[4] in the limit $J_{-} \rightarrow \infty$). The time evolution of the  state is given by the stochastic Schrodinger equation,
\begin{equation}
    i\frac{d}{d t} \ket{\psi^{\eta}_k(t)}= \mathcal{H}^{\eta}_k(t) \ket{\psi^{\eta}_k(t)},
\end{equation}
where the corresponding density matrix obtained for the single realization of noise, $\hat{\rho}^{\eta}_k (t)=\ket{\Psi^{\eta}_k(t)} \bra{\Psi^{\eta}_k(t)}$, is  averaged over the noise to yield  $\overline{\hat{\rho}^\eta_k(t)}$. With the initial conditions, $c^{\eta}_{1k}(-\infty)=1$ and $c^{\eta}_{2k}(-\infty)=0$, 
we set up the Bloch equation for the population inversion ($\rho^{\eta}_k$) after considering the transformation,  $\tau=2 \Delta_k t$ and $v_{_{\mathrm{LZ}}}=v \sin{\frac{k}{2}}/ \Delta^{2}_k$ (where, $\Delta_k=2 J_{+} \cos{\frac{k}{2}}$), 
\begin{eqnarray}
\frac{d}{d\tau}\rho^{\eta}_k(\tau) = -\frac{1}{2} \int^{\tau}_{-\infty} d\,\tau_1 e^{i\, \int^{\tau}_{\tau_1} d \tau_2\, \frac{ v_{_{\mathrm{LZ}}} \tau_2}{2}}
\,\rho^{\eta}_k(\tau_1)\notag{}\\
-\frac{1}{2\,J^2_{+}}  \int^{\tau}_{-\infty} d\,\tau_1 e^{i\, \int^{\tau}_{\tau_1} d\tau_2\,  \frac{ v_{_{\mathrm{LZ}}} \tau_2}{2}}
  \eta(\tau) \eta(\tau_1) \,\rho^{\eta}_k(\tau_1) + h.c. .
\end{eqnarray}
Following the approaches similar to that given in~\cite{Sinitsyn2003, Fai2013}, the reduced master equation is solved by averaging the Bloch equation  
over the fast Gaussian noise to yield in the long time limit, $$\overline{\rho_k}(\infty)= e^{-2\pi\, \eta_0^2/(2\,v_{_{\mathrm{LZ}}}\, J^2_{+})}\Big(2\,e^{-\pi/(2\, v_{_{\mathrm{LZ}}})}-1\Big),$$ and the  noise averaged excitation probability is given by, 
\begin{equation}
p_k= \frac{1}{2}\left[1+ e^{-\chi \frac{ \cos^2{k/2}}{\sin{k/2}}} (2 e^{-2\pi \tau_Q   \frac{ \cos^2{k/2}}{\sin{k/2}}}-1) \right]
\label{eq:pk}
\end{equation}
where $\chi=4 \pi \eta_0^2 \tau_Q$ and we have set $J_{+}=1$. Notice that the non-adiabatic region responsible for increased excitations opens up around $k=0$  (as can be seen from the Fig.~[\ref{fig:pk0}] of the excitation probability Eq.~[\ref{eq:pk}]) instead of  $k=\pi/2$ region which is the case for the  TFQIM or XY spin-chian case with transverse drive protocol~\cite{PhysRevLett.117.080402,PhysRevB.95.224303,PhysRevB.104.064313}.
For $\eta_0=0$, one  retrieves the excitation probability as given by the LZ formula for the noiseless drive case~\cite{PhysRevB.79.224408, PhysRevB.81.054306},
\begin{equation}
    p^{0}_{k}=e^{-2\pi \tau_Q \frac{ \cos^2{k/2}}{\sin{k/2}}}.
\end{equation}

 In the thermodynamic limit ($N\rightarrow \infty$) and in the long time limit ($t\rightarrow \infty$), the noise averaged decohered density matrix can be written as, $\overline{\rho}_D=\Pi_{k>0} \overline{\rho}^k_D$, with  $\overline{\rho}^k_D$ given by,
 \begin{equation}
    \overline{\rho}^k_D= p_k \ket{\Bar{e}_{1k}} \bra{\Bar{e}_{1k}}+(1-p_k )\ket{\Bar{e}_{2k}} \bra{\Bar{e}_{2k}},
\label{eq:rho_D}
\end{equation}
which  characterizes the mixed state i.e., the noise averaged coherences of the density matrix, $\overline{\rho^k_{12}}$ and  $\overline{\rho^k_{21}}$ involving  terms like
$\ket{\Bar{e}_{1k}} \bra{\bar{e}_{2k}}$  and $\ket{\Bar{e}_{2k}} \bra{\Bar{e}_{1k}}$ are highly fluctuating and vanishes in the long time limit.

\section{Defect Density and Residual Energy}\label{DDRE}
For the noiseless drive through the QCP,  the defect density is obtained by integrating the excitation probability i.e., $ n_0=\frac{1}{\pi}\int^{\pi}_{0} d k  \, p^{0}_k$. 
  For the slow sweep regime, the dominant contribution for the defect generation comes from the non-adiabatic region around the critical point at $k=\pi$ (see the inset of  Fig.~[\ref{fig:pk0}]). Therefore,
 the exponent of $p_k^0$ acquires a quadratic form and the integral yields
$n_0\approx1/  \pi\sqrt{2 \tau_Q}$, implying that
  the defect density follows the KZ scaling behavior,
  $n_0 \sim \tau^{-\beta}_Q, $ with $\beta=d \nu/(1+z \nu)=1/2$ (where dimension $d=1$  and the critical exponents are $\nu=z=1$) ~\cite{PhysRevB.81.054306}. 

\begin{figure}
    \centering
    \includegraphics[width=0.49\columnwidth,height=4.5cm]{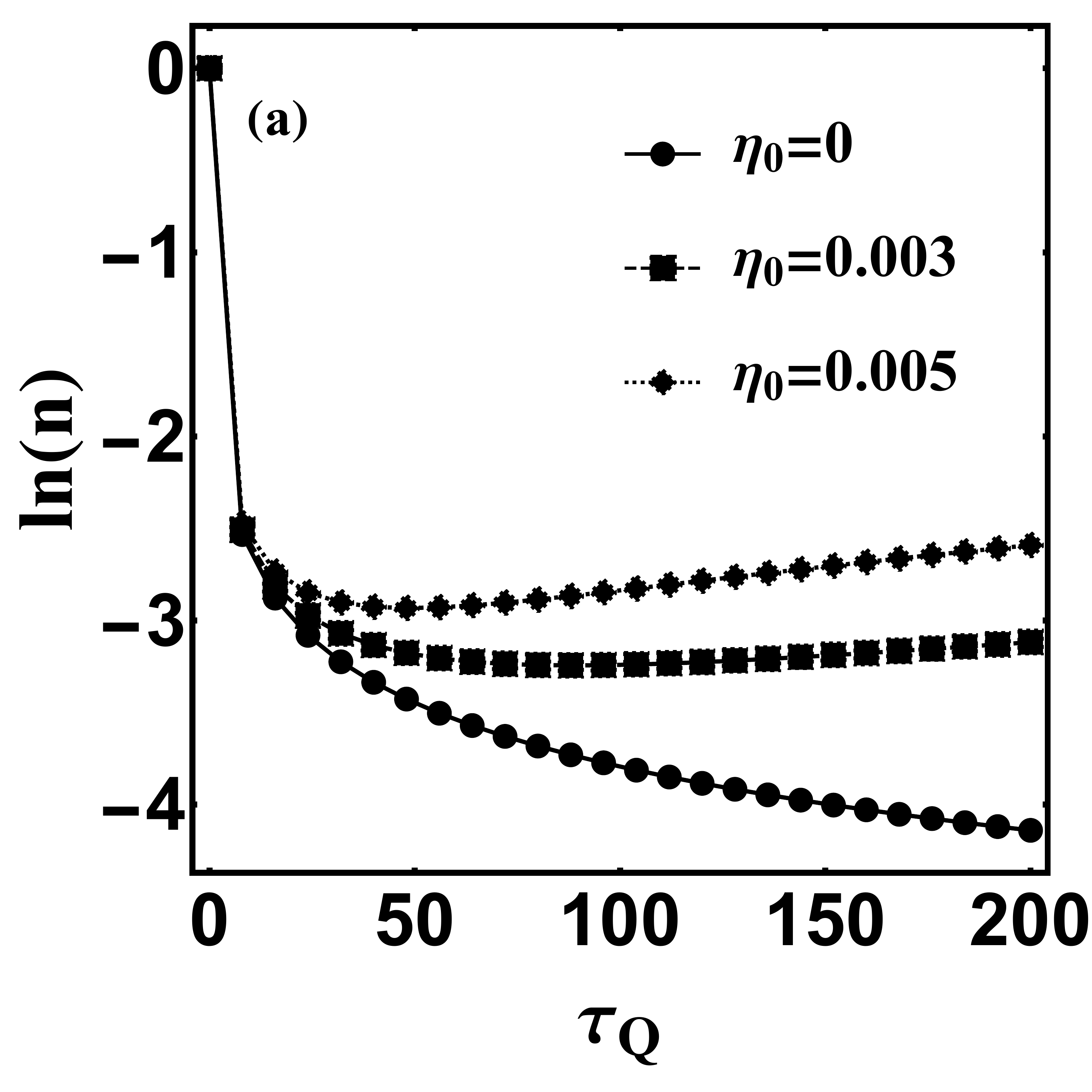}
    \includegraphics[width=0.49\columnwidth,height=4.45cm]{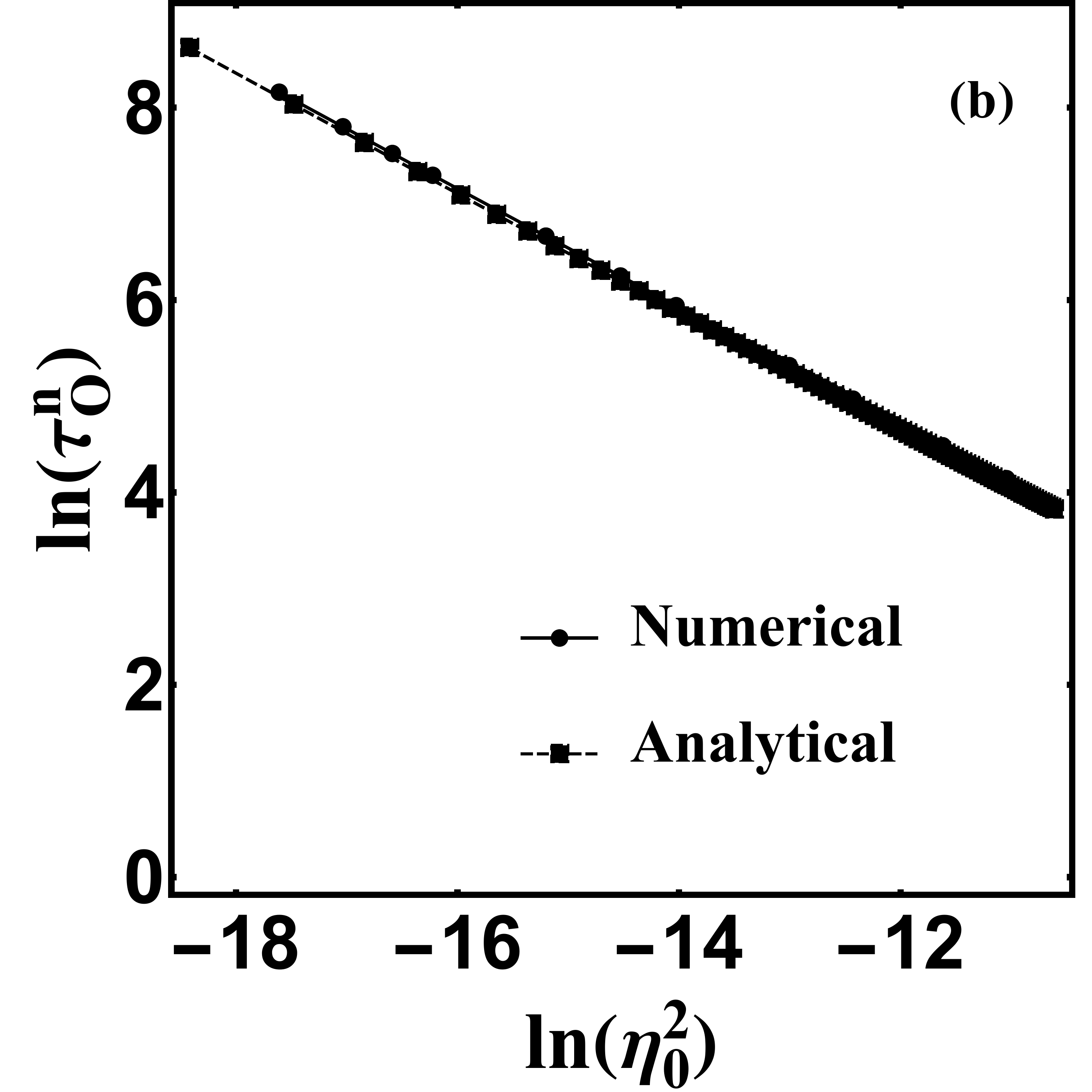}
    \includegraphics[width=0.49\columnwidth,height=4.5cm]{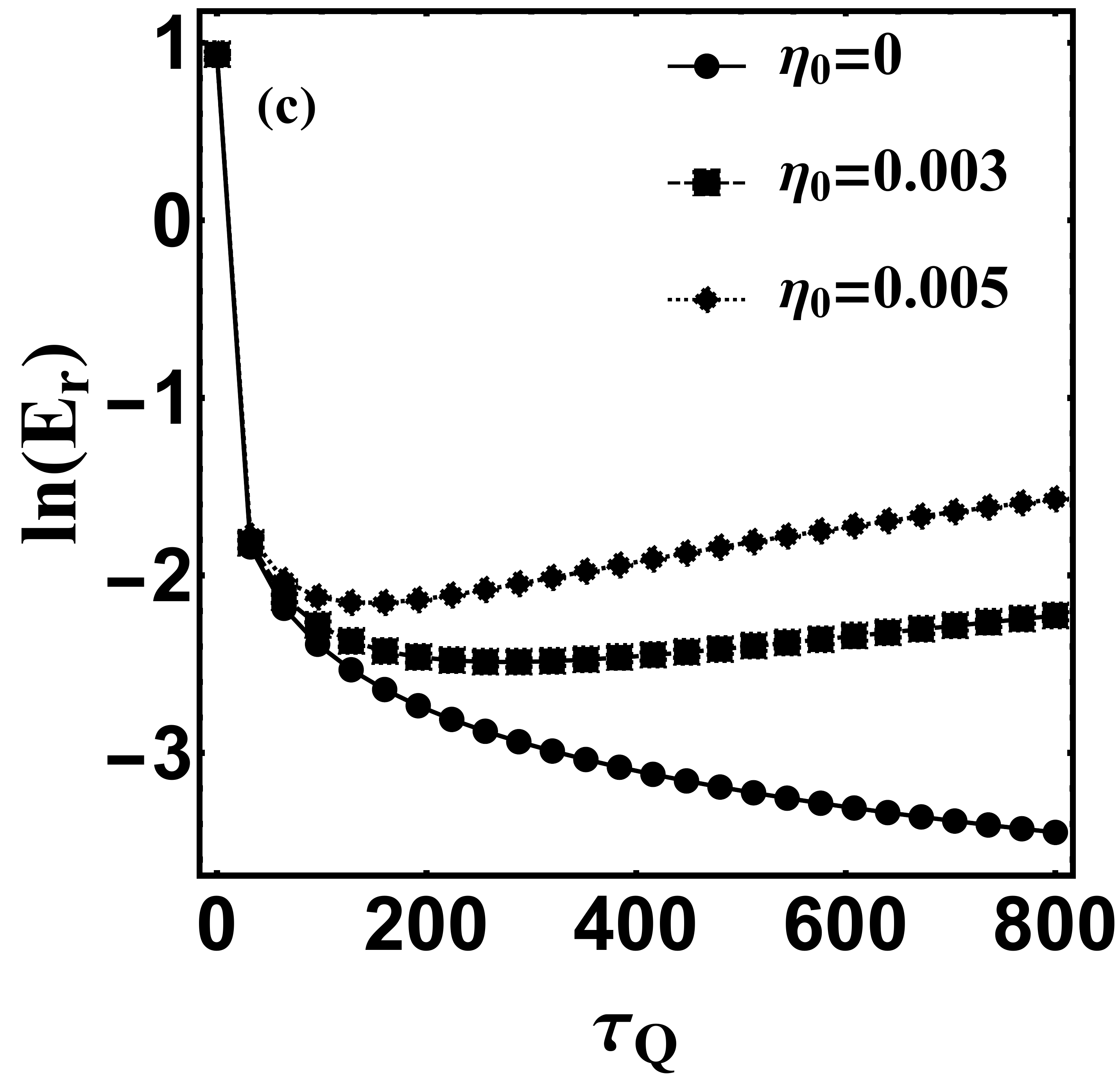}
    \includegraphics[width=0.49\columnwidth,height=4.5cm]{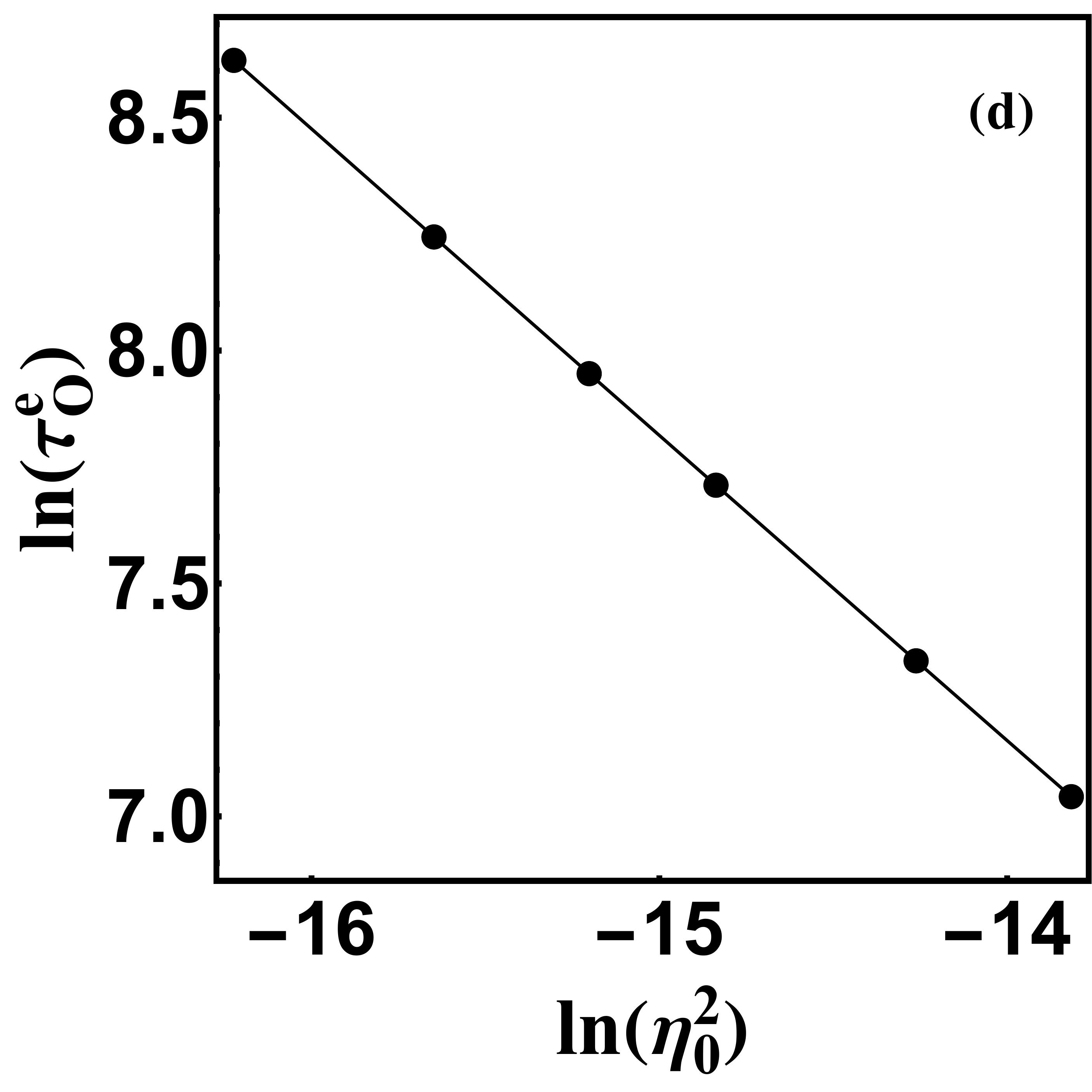}
    \caption{(a) Defect density $n$ as a function of quench time shows AKZ behavior for non-zero noise amplitude. (b)  The log of  optimal quench time extracted from the defect density as a function of $\log \eta_0^2$ exhibits strong agreement between the one obtained via  the numerical approach and  by using the approximate analytical result for $\tau^n_O$. The best fit line exhibits deviation from the usual $\eta_0^{-4/3}$ optimal quench time scaling behavior. 
    		  (c) The residual Energy per  site,  $E_{r}$, as a function of quench time also exhibits AKZ behavior  for non-zero noise amplitude.  (d) The quench time, $\tau_O^e$, to minimize the   residual energy exhibits, $\eta_0^{-4/3}$,       power-law scaling behaviour with the noise strength.}
    \label{fig:n}
\end{figure}

The defect density for the noisy drive, 
\begin{equation}
n=\frac{1}{\pi}\int^{\pi}_{0} p_k d k,
\end{equation}
with $p_k$ given by Eq. \ref{eq:pk}
is evaluated numerically. Unlike the noiseless case ($\eta_0=0$), the defect density for noisy drive exhibits AKZ behavior i.e.,  slower drive results in increased defect creation with the increase of noise strength as seen from  Fig.[\ref{fig:n}].   The behavior of the defect density can be split in to two asymptotic regimes depending on the noise strength  $\eta_0$. In either case we will focus on the slow sweep scenarios ($\tau_Q\gg 1$). For $\chi \ll 1 $, the defect density can be approximated as 
\begin{equation}\label{eq:n_weak}
n\approx  \frac{1}{\pi \sqrt{2\tau_Q }}-\frac{\chi}{\pi} \log(\chi),
\end{equation}
where the first term is the usual KZ behavior. 
 The second term  exhibits AKZ behavior which is present due to the contribution from the $k\approx 0$ region of the excitation probability (see  Fig.~[\ref{fig:pk0}]), in particular from the term,  $\int dk [1-\exp(-\chi\cos^2k/2 /\sin(k/2))]$.  
It is multiplied by  $\log \chi$ and hence is more dominant than the  AKZ terms obtained in earlier studies~\cite{PhysRevLett.117.080402, PhysRevB.95.224303, PhysRevB.104.064313}. The quench time which minimizes the defect density, Eq.~\ref{eq:n_weak} , also referred to as the optimal quench time can be  estimated to be,
\begin{equation}
\tau^{n}_O\approx \big[ 8\sqrt{2}\pi\eta_0^2\big(\log\frac{1}{4\pi\eta_0^2 \tau_0}-1\big)\big]^{-2/3},
\label{eq:n_optimal}
\end{equation}
where $\tau_0 = \big[ 8\sqrt{2}\pi\eta_0^2(\log\frac{1}{4\pi\eta_0^2 }-1)\big]^{-2/3}.$
In the second limit, $\chi \gg 1 $,  the defect density can be approximated as 
\begin{equation}
n\approx \frac{1}{2}\Big(1-\frac{1}{\sqrt{\pi\chi} }\Big) + \frac{1}{\pi \sqrt{2\pi \tau_Q   + \chi}},
\end{equation}
thus for very slow sweep rates the noise tends to completely scramble the initial state i.e, $n\rightarrow \frac{1}{2}$.

The effect of the quench through the critical points is also imprinted on the residual energy. The residual energy is the excess energy relative to  the ground state of the Hamiltonian at the end of the drive protocol.
The expression for  the residual energy per site in the units of $J_-$ is given by~\cite{refId0}, 
\begin{equation}
E_r =\text{lim}_{t,N \rightarrow \infty}\frac{\text{tr}\,[\Bar{\rho}^{D}\,H]-E_0}{N |J_{-}(t)|} =\frac{4}{\pi}\int^{\pi}_{0}dk~ p_k \sin{\frac{k}{2}},
\end{equation}
where $\text{tr}\,[\Bar{\rho}^{D}\,H]$ is the expectation value of the Hamiltonian with respect to the final decohered state, $\bar{\rho}^{D}=\Pi_{k>0} \overline{\rho}^k_D $ and $E_0$ is the total energy of  the ground state. The numerically evaluated residual energy shows the signature of AKZ behavior for slower drives, as seen from Fig.[\ref{fig:n}].  Analogous to the derivation of the limiting behavior of the defect density, in the limit $\chi \gg 1 $, $E_r \approx 4 n$.  In the opposite limit, $\chi  \ll 1 $, $E_r\approx 2\sqrt{2}/\pi \sqrt{ \tau_Q}  + \chi$, i.e., the KZ term  has the same form as that for $n$, except now it is multiplied by a factor of $4$, while the AKZ term  has the conventional 
linear in $\chi$ dependence. Consequently,  the quench time which minimizes the residual energy has a conventional scaling behavior  as $\eta_0^{-4/3}$~\cite{PhysRevLett.117.080402}, which is also  confirmed by the Fig.~\ref{fig:n}.

\begin{figure}
	\centering
	\includegraphics[width=\columnwidth,height=8cm]{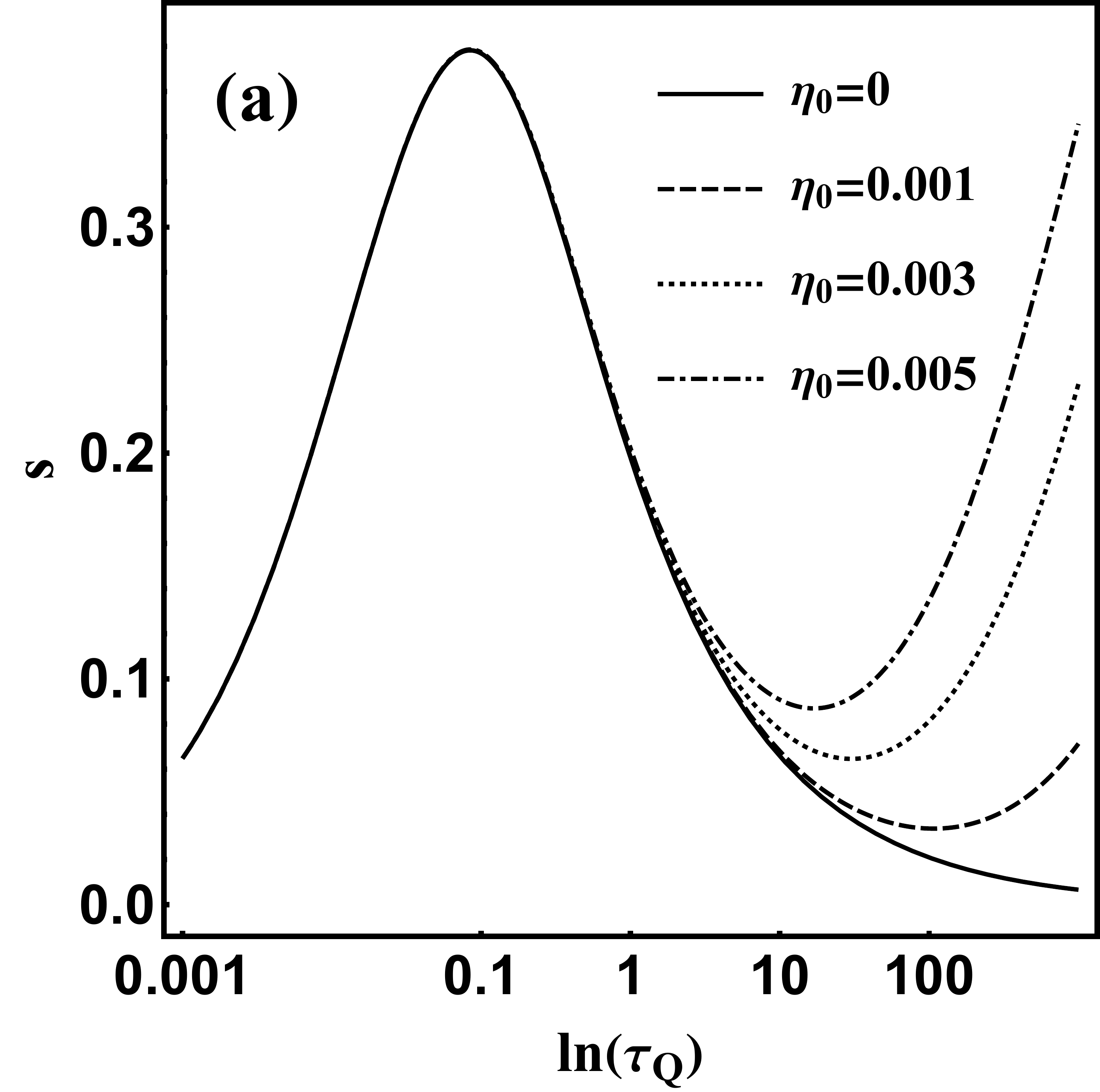}\vspace{0.85mm}
	\includegraphics[width=0.49\columnwidth,height=4cm]{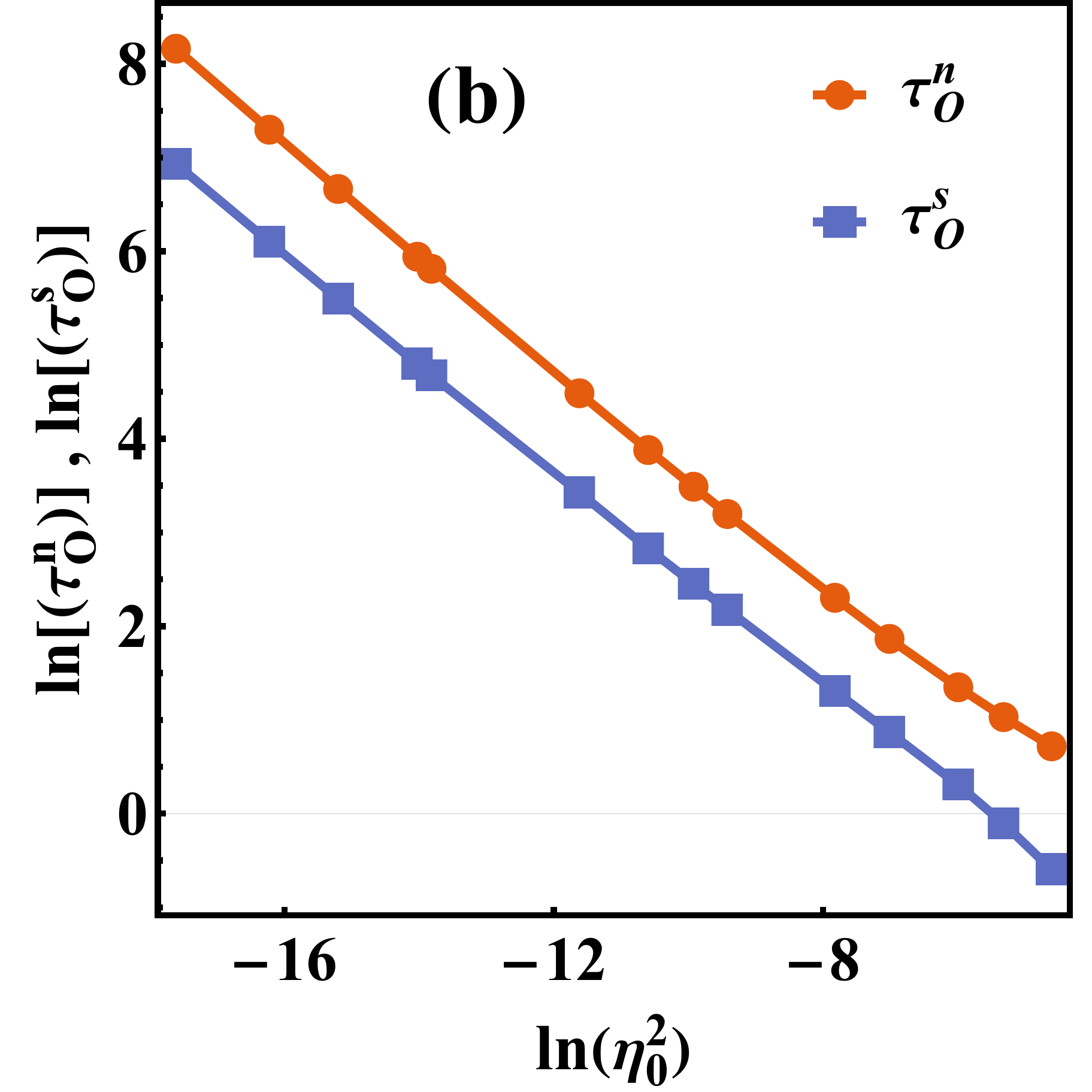}
	\includegraphics[width=0.49\columnwidth,height=4cm]{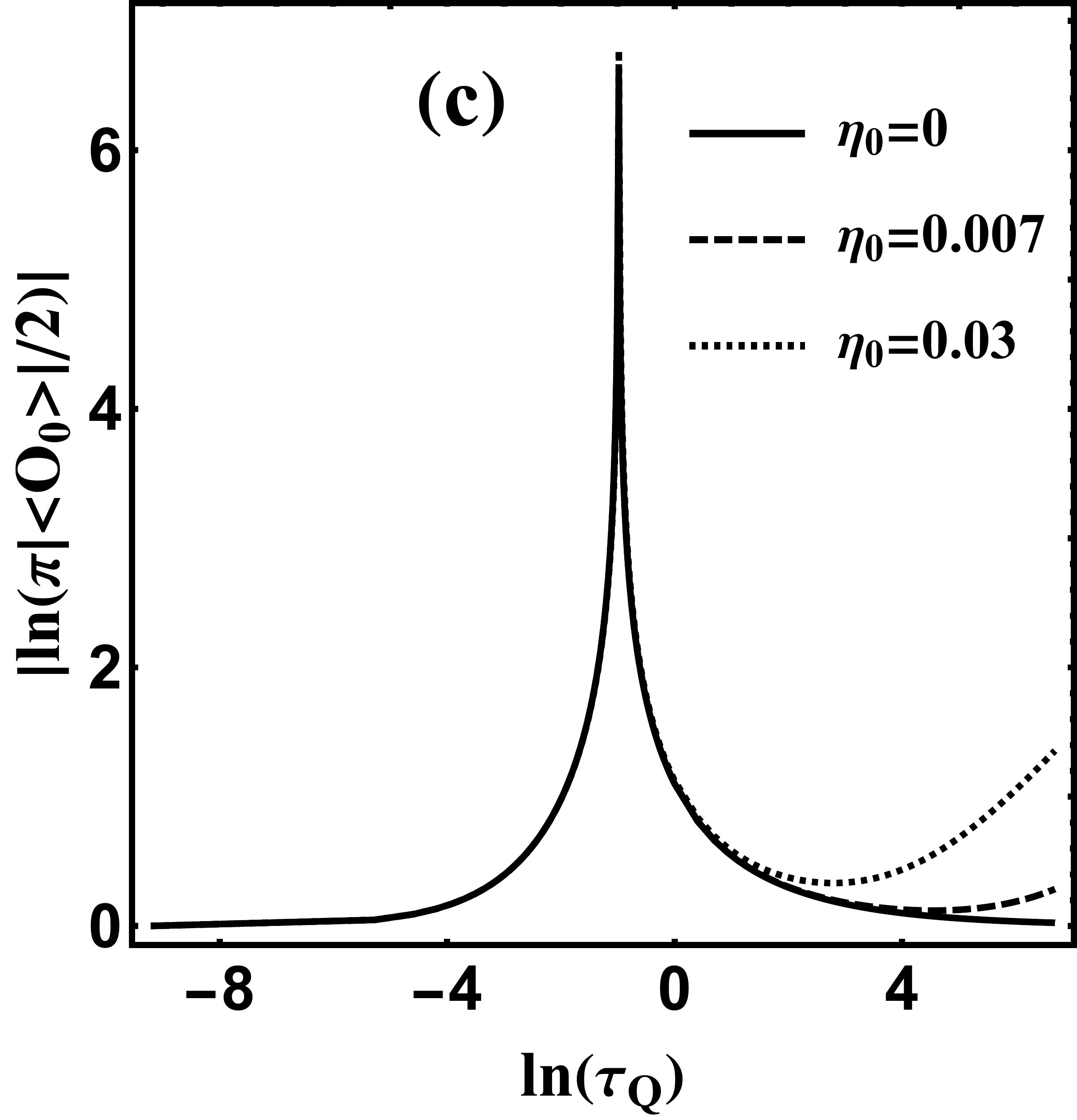}
	\caption{ (a) Entropy density as a function of the quench time exhibits AKZ behavior   at slower quench times (the entropy density exhibits an upturn) due to the noise induced increased entropy production.  
		(b)The log-log plot of the optimal quench time, $\tau^{n}_{O}$  and $\tau^{s}_{O}$ (which is extracted from the minimized noisy entropy density) has been plotted with respect to the noise strength ($\log \eta_{0}^2$) which clearly shows that 
		$\tau^{n}_{O}$ and $\tau^{s}_{O}$ are proportional.  (c) The behavior of   $|\log (\pi |\langle O_{0}\rangle |/2)|$ is qualitatively similar to that of the entropy density.}
	\label{fig:entropy1}
\end{figure}

\section{Entropy Density}\label{EDensity}
A finite non-zero entropy density is a signature of mixed state which can  result from either the internal or external decoherence during the drive through the critical points. The von Neumann entropy density in the final decohered state $\bar{\rho}_D$ is given by,
\begin{equation}
s=-\frac{1}{\pi}\int^{\pi}_0 d k\,[p_k \,\text{ln}\, p_k +(1-p_k)\ \text{ln}\, (1-p_k)],
\label{eq:entropy}
\end{equation}
where the numerically integrated entropy density as a function of the quench time has been plotted in Fig.~[\ref{fig:entropy1}]. 
In the noiseless drive scenario consistent with the KZ picture, the entropy density is small in both the fast ($\tau_Q \ll 1$) and the slow sweep regime ($\tau_Q \gg 1$). Furthermore, there is an intermediate $\tau_Q$-regime between the fast and the slow sweep regimes where the entropy density maximizes. For the noisy drive case, we observe (see Fig.~[\ref{fig:entropy1}]) that for slow sweep regime the entropy density shows an upturn for  quench times of the order of the noise strength dependent optimal quench time.
 In the $\chi \gg 1$ limit, the entropy density can be approximated by, $ s\approx \text{ln}2-\frac{1}{4\eta_0\sqrt{2\tau_{Q}}}$,
	which approaches the maximum entropy state ($s\rightarrow \text{ln}\,2$) for sufficiently slow sweeps, i.e.,  the noise randomizes the system completely.
The noisy drive is characterized by two quench time scales, first, the intermediate quench time scale at which the entropy density locally maximizes and the second quench time scale/regime characterized by the optimal quench time ($\tau^e_O$) beyond which results in increased entropy production due to the presence of the noise, i.e.,  AKZ behavior~\cite{ PhysRevB.104.064313}. As can be seen from the Fig.~[\ref{fig:entropy1}], the optimal quench time scale ($\tau^e_O$) related to entropy density exhibits similar power law dependence with the noise strength as shown by the optimal quench time ($\tau^n_O$) associated with the defect density.

Interestingly,  this behavior  has qualitative similarities  with the behavior of $ |\log (\pi |\langle O_{0}\rangle |/2)|$ (where $\langle O_{0}\rangle$ is the two-point
Majorana correlator at zero separation as defined in Sec.~\ref{StrCorr})
 as can be seen Fig.~[\ref{fig:entropy1}]
and   can also be associated with the crossover behavior of   $\langle |O_{0}|\rangle$~\cite{PhysRevB.78.045101}. 
For $\tau_Q\ll 1$, the excitation probability $p_k$ for the  noiseless drive tends to $1$ for all the $k$-modes, therefore the entropy density will be small and one notices that  $ |\log (\pi |\langle O_{0}\rangle |/2)|$ also vanishes. On the other hand, for  the $\tau_Q\gg 1$ and the noiseless case, the excitation probability $p_k$ will be close to zero for the all the modes except near the non-adiabatic region around $k=\pi$ point,  where the entropy density and at the same time  $ |\log (\pi |\langle O_{0}\rangle |/2)|$ obtains vanishingly small contribution.
Moreover, in the noisy regime similar to the entropy,  $ |\log (\pi |\langle O_{0}\rangle |/2)|$, exhibits upturn for the slow sweep regimes.

\section{String Correlations} \label{StrCorr}
The transition at $J_1=J_2$ across the two gapped fermionic state corresponds to the change of topological order rather than the change of symmtery~\cite{PhysRevLett.98.087204, KITAEV20062}. It turns out that the two-point spin-spin correlator vanishes for larger than one separation  in either regimes~\cite{PhysRevLett.98.247201}, while in the ``ordered phase''  the highly non-local string operator correlations acquire non-vanishing values~\cite{PhysRevLett.98.087204, Chen_2008, Chitov_2017, PhysRevB.100.104428, Bahovadinov_2019}, moreover, it was shown by performing the spin-duality transformation that one can map the $1$-D Kitaev chain to $1$-D Ising chain.  The local operators underlying order to disorder transition in the dual space are highly non-local in terms of the original spin operators. These are referred to as the hidden string order parameters (SOP)~\cite{PhysRevLett.98.087204}.

We begin by calculating the two-point Majorana operators of the type $O_r=ib_n a_{n+r}$, which in the spin language corresponds to the string operators. Our aim is to understand the long  range order for slower sweeps in the final decohered/mixed state both for the noiseless and the noisy scenarios.
We will next calculate the spin-spin correlations in the dual space, these are expressed in terms of the product of two-point  Majorana correlators.
\subsection{Two-point Majorana correlator}
The two-point Majorana correlator denoted by $O_r=ib_n a_{n+r}$ 
has the following expression in  terms of the fermion operators $a_k$ and $b_k$ ,
\begin{equation}
    O_r =\frac{4 i}{N}\sum_{k=0}^{\pi}(b_k^\dagger a_k e^{ikr}+b_k a_k^\dagger e^{-ikr}+b_k a_k  e^{-ikr}+b_k^\dagger a_k^\dagger  e^{ikr}).
\end{equation}
In the limit $J_{-}\rightarrow\infty$,  keeping $J_{+}$ fixed,  the eigenvectors  of the Hamiltonian  Eq.~[\ref{eq:H_matrix1}] are given by, 
\begin{eqnarray}
\ket{\Bar{e}_1}&=&\frac{1}{\sqrt{2}}(a^\dagger_ke^{\frac{-ik}{2}}+b^\dagger_k)\ket{0}{}\nonumber\\
\ket{\Bar{e}_2}&=&\frac{1}{\sqrt{2}}(-a^\dagger_ke^{\frac{-ik}{2}}+b^\dagger_k)\ket{0}
\end{eqnarray}
where the eigenvalues are, $\Bar{E}_{1k,2k}=\pm 2J_{-}\sin{\frac{k}{2}}$ for the the excited state $\ket{\Bar{e}_{1k}}$ and the ground state $\ket{\Bar{e}_{2k}}$, respectively.  
The final decohered state has the  
following form 
$\overline{\rho}_D=\Pi_{k>0} \overline{\rho}^k_D$, where 
$\overline{\rho}^{k}_{D}$  is the 
density matrix corresponding to the 
$k^{\text{th}}$-mode given by,
$ \overline{\rho}^k_D= p_k \ket{\Bar{e}_{1k}} \bra{\Bar{e}_{1k}}+(1-p_k )\ket{\Bar{e}_{2k}} \bra{\Bar{e}_{2k}}$, which is a mixed state.

The expectation value of $O_r$,  with respect to the final decohered state i.e., $\langle O_r \rangle =\text{tr} (\overline{\rho}_D O_r)$  is given by,
\begin{equation}
    \langle O_r \rangle =\sum_{k=0}^{\pi}p_k\bra{\Bar{e}_{1k}}O_r\ket{\Bar{e}_{1k}}+(1-p_k)\bra{\Bar{e}_{2k}}O_r\ket{\Bar{e}_{2k}}.
\end{equation}
Using the result, 
 $\bra{\Bar{e}_{1k}}b_k a_k^\dagger\ket{\Bar{e}_{1k}} =-\frac{1}{2}e^{\frac{ik}{2}}$, 
 and  $\bra{\Bar{e}_{2k}}b_k a_k^\dagger\ket{\Bar{e}_{2k}} =\frac{1}{2}e^{\frac{ik}{2}}$, 
the expectation values of  $\langle O_r \rangle$ yields the following expression
in the thermodynamic limit ($N\rightarrow\infty$)
\begin{equation}
    \langle O_r \rangle=\frac{2}{\pi}\int^{\pi/2}_{0}\,d k(1-2 \,p_k)\sin{k(2r-1)}.
    \label{eq:majorana_corr1}
\end{equation}
Following similar approach we find, 
\begin{equation}
\langle i a_n a_{n+r}\rangle =i\int_{0}^{\pi}dk \cos{kr}=i\delta_{r,0},
\end{equation}
 and $\langle i b_n b_{n+r}\rangle =-i\delta_{r,0}$.
\begin{figure}
    \centering
    \includegraphics[width=0.49\columnwidth]{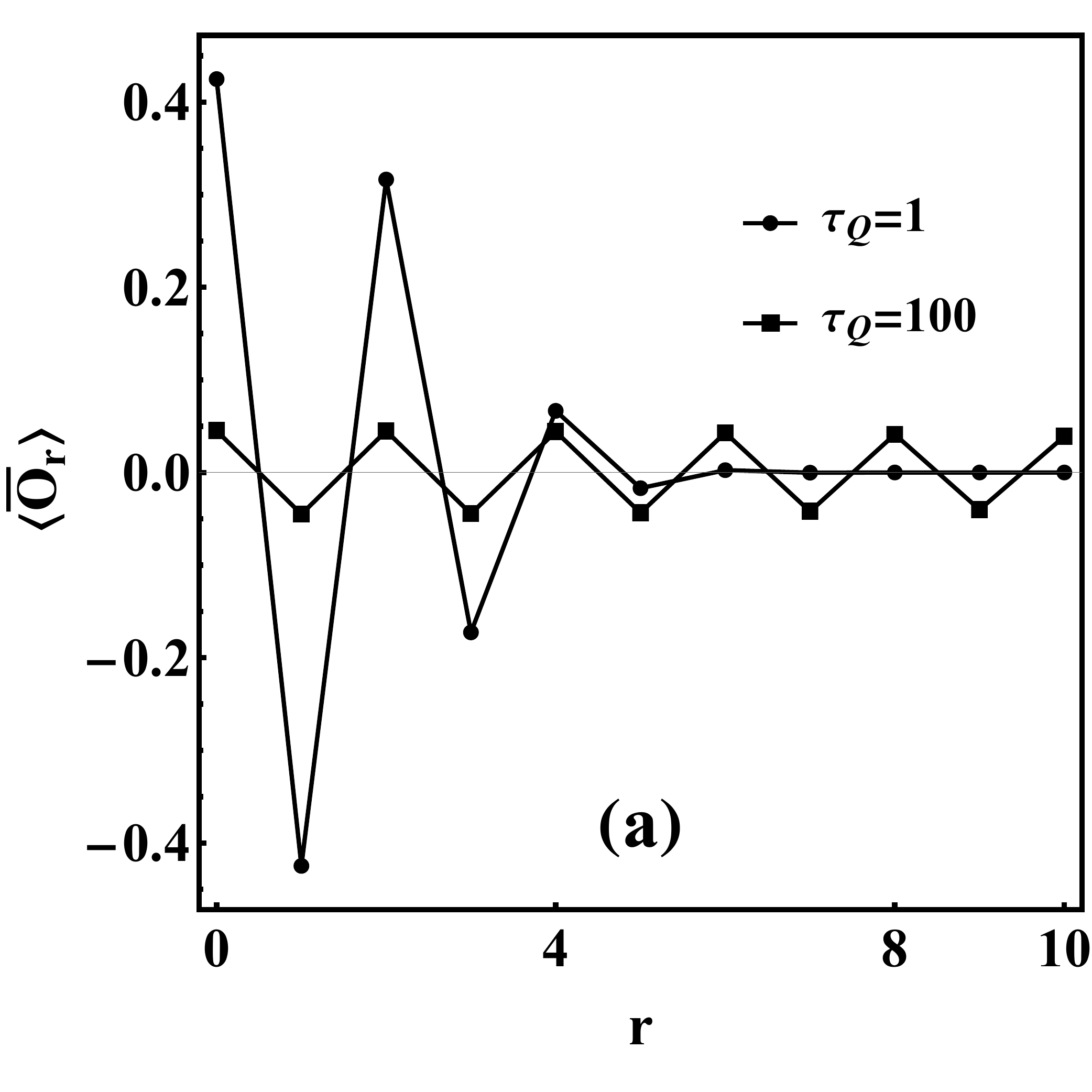}
    \includegraphics[width=0.49\columnwidth]{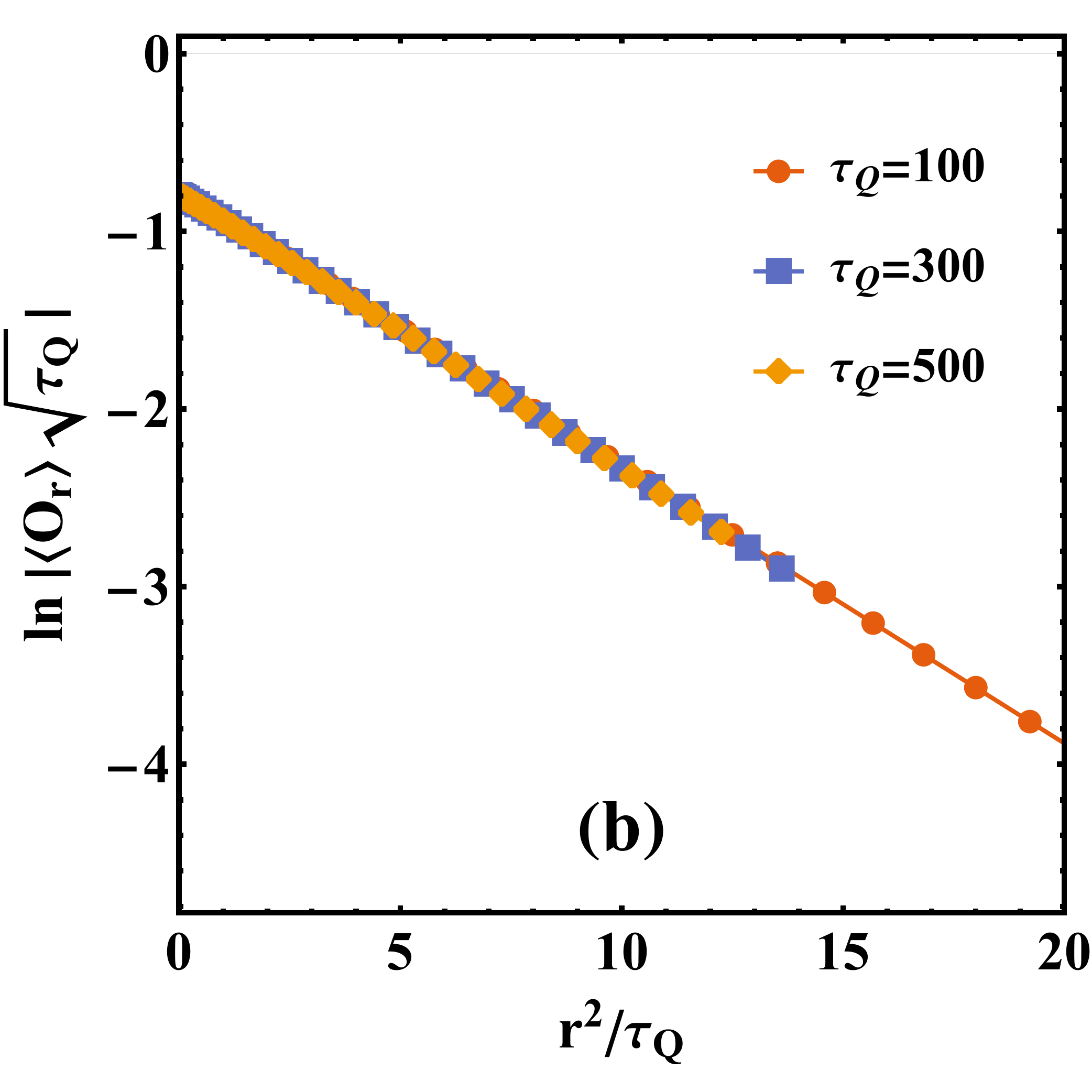}
    \caption{ (a) $\langle \bar{O}_r \rangle$ as a function of  $r$ exhibits damped oscillatory behavior in the absence of noise. For the slower sweep speeds the  spatial extension of the two-point Majorana correlator is long ranged. (b) As observed from the rescaled figure, the correlation length associated with the two point Majorana correlator $\langle \Bar{O}_r \rangle$ exhibits the following scaling behaviour, $ e^{-(2r-1)^2/8\pi\tau_Q}/\sqrt{ \tau_Q}$, with  the correlation length  proportional to $\sqrt{\tau_{Q}}$.}
    \label{fig:majorana_corr}
\end{figure}
\begin{figure}
    \centering
    \includegraphics[width=0.49\columnwidth]{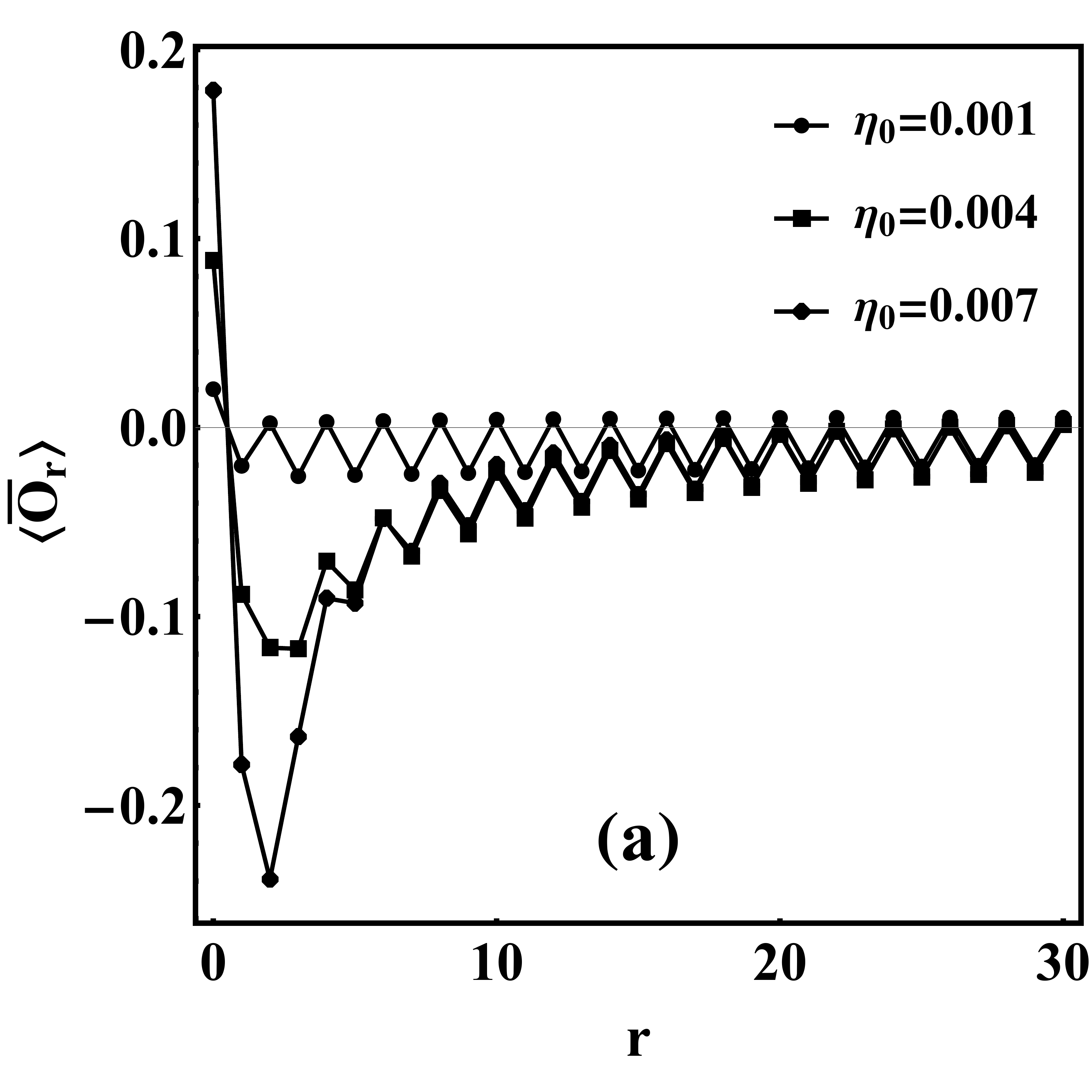}
    \includegraphics[width=0.49\columnwidth]{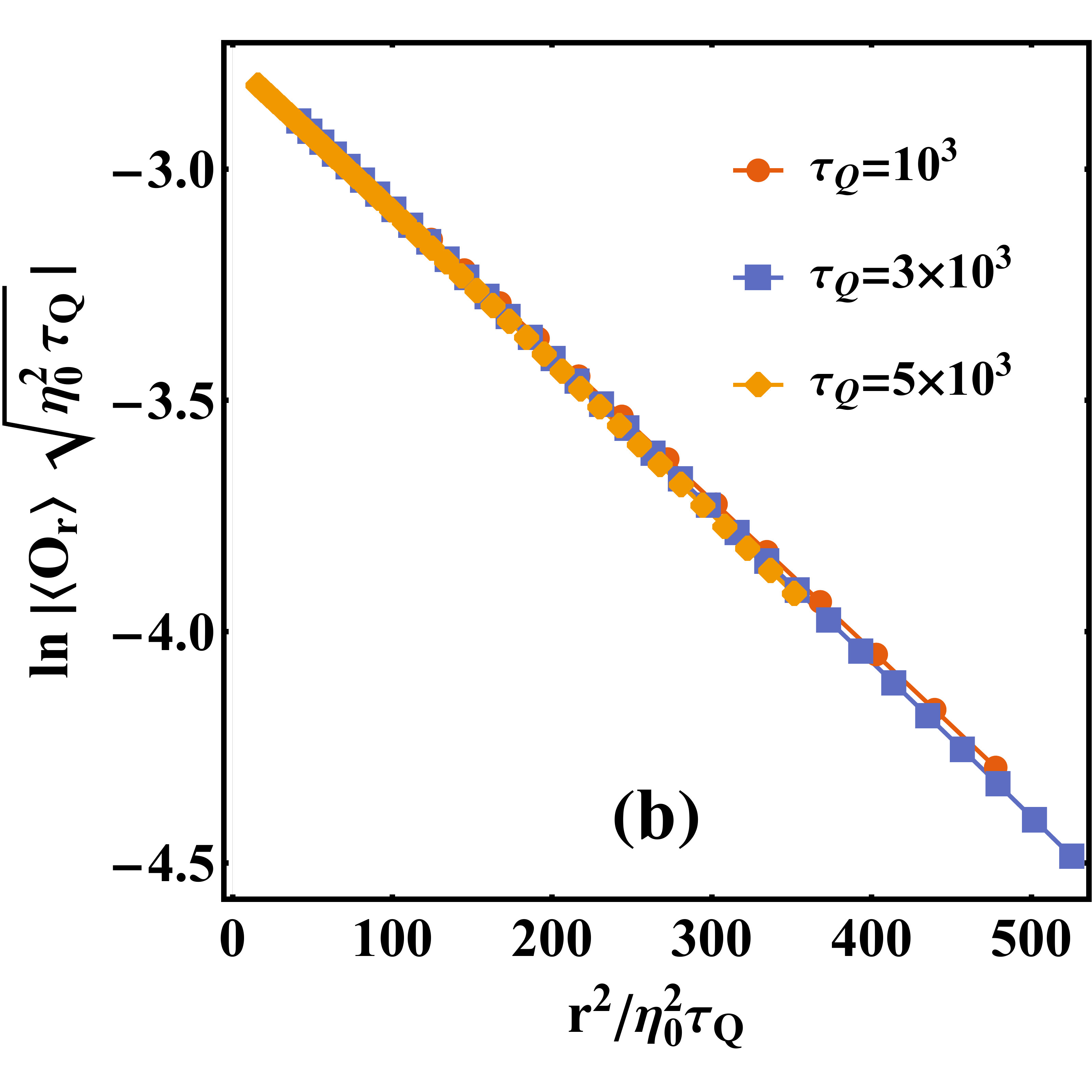}
    \caption{(a). In the regime  $\chi=4\pi\eta_0^2 \tau_Q\ll 1$ with $\tau_Q=1000$, or the effectively weak noise regime,  $\langle \bar{O}_r \rangle$  vs. $r$-separation continues to exhibit  damped oscillatory behavior, however there is a downward shift as predicted by Eq.~\ref{Eq:O_rweak}. (b) In the $\chi \gg 1$ regime (here $\eta_0 =0.009$) $\langle O_r \rangle$ exhibits exponential scaling  with respect to $r^2/ \eta_0^2 \tau_Q$    in confirmation with Eq.~\ref{Eq:O_rstrong}.} 
    \label{fig:majorana_corr_noisy}
\end{figure}
In the noiseless regime, $\langle O_0 \rangle$  for 
$\tau_Q\ll 1$  and $\tau_Q\gg 1$ differs simply by a sign. This is because $p_k\rightarrow 1$ for $\tau_Q\ll 1$
while it vanishes for the slow sweep regime. Similar observation was made in Ref.~\cite{PhysRevB.78.045101} although over there
the quenching scheme involved changing $J_1$, moreover the  expression for $\langle O_r \rangle$  is different from that given in Eq. \ref{eq:majorana_corr1}.
We note that the first term corresponds to the defect free part of the expectation value of $\langle O_r \rangle$ and yields a power law decay with distance, $\frac{1}{\pi(  r-1/2)}$.
The second term on the other hand accounts for the measure of spatial extension of the defect correlations in the final decohered state and for the slow sweep and the noiseless scenario is given by, $(-1)^r\sqrt{2/\pi^2 \tau_Q}~ e^{-(2r-1)^2/8\pi\tau_Q }. $
 The plot of  $\langle \bar{O}_r \rangle=\langle O_r \rangle-(\text{defect free part})$, shown in Fig.~[\ref{fig:majorana_corr}] for the noiseless case, shows the spatial extension of the defect correlation which clearly has oscillatory damped behavior (long ranged order) in the slow sweep regime.
For the noisy drive scenario, there is an increment in the defect generation for the slow sweep regime therefore we expect  shift of $\langle \Bar{O}_r \rangle$ towards negative values as can also be observed from the Fig.~{\ref{fig:majorana_corr_noisy}. In particular, for $\chi \ll 1$ and $r$ such that  $\chi (2 r -1) \leq 1$,  $\langle \Bar{O}_r \rangle$ is given by
\begin{equation}
\langle \bar{O}_r \rangle \approx -\chi[1 -\frac{\chi (4 r-2)}{\pi}] + \sqrt{\frac{2}{\pi^2  \tau_Q}}e^{-\frac{(2r-1)^2}{8\pi\tau_Q} +i \pi r}.\label{Eq:O_rweak}
\end{equation}
For the second scenario, $\chi \gg 1 $ and therefore $p_k\approx 1/2$, the power-law decay of $\langle O_r \rangle$  present in the weak noise case is completely absent and  instead
replaced by  exponential decay terms as can also be verified from rescaled plot in Fig.~[\ref{fig:majorana_corr_noisy}]  and has  the following form
\begin{equation}
\langle O_r \rangle \approx  -(-1)^r\frac{e^{-(2r-1)^2/4\chi }}{2 \pi  \sqrt{\chi}}.\label{Eq:O_rstrong}
\end{equation}
We note that  the correlation length  attributed to the exponential decay is reduced by a factor of $\sqrt{ \chi/2 \pi \tau_Q}$ as compared to the noiseless case. 
\subsection{Hidden String Correlator}
\begin{figure}
    \centering
    \includegraphics[width=0.49\columnwidth]{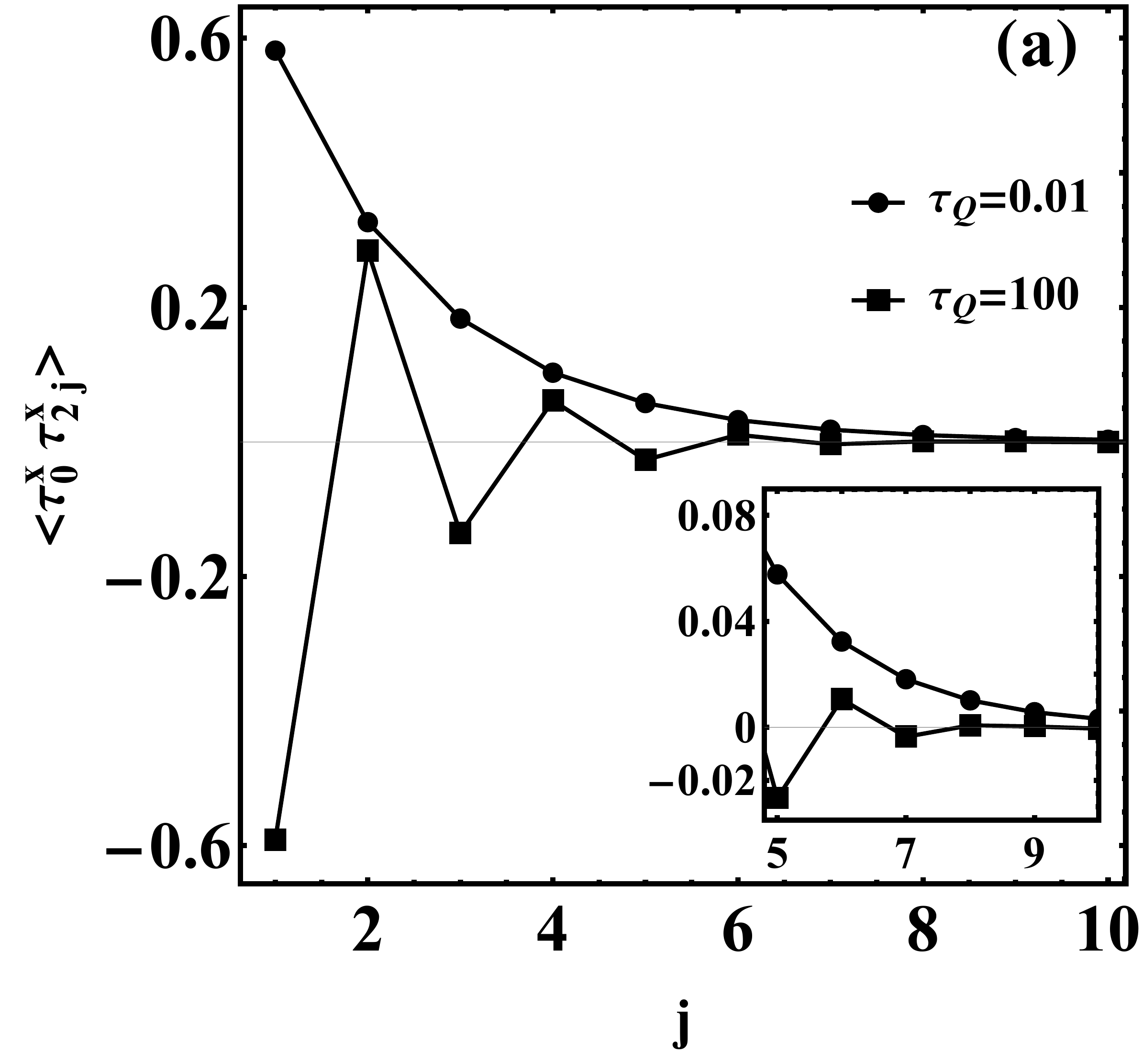}
    \includegraphics[width=0.49\columnwidth]{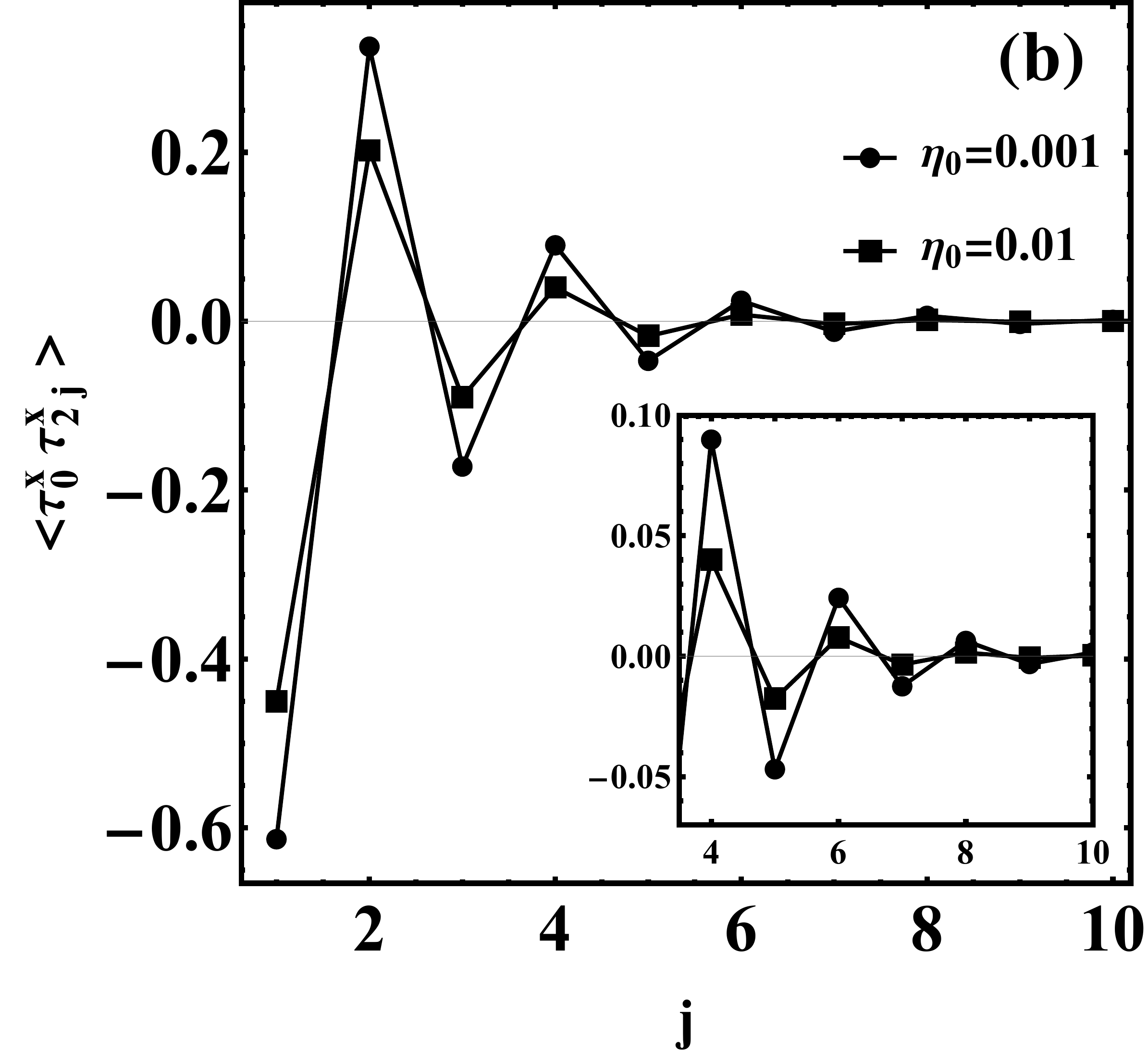}
    \includegraphics[width=0.49\columnwidth]{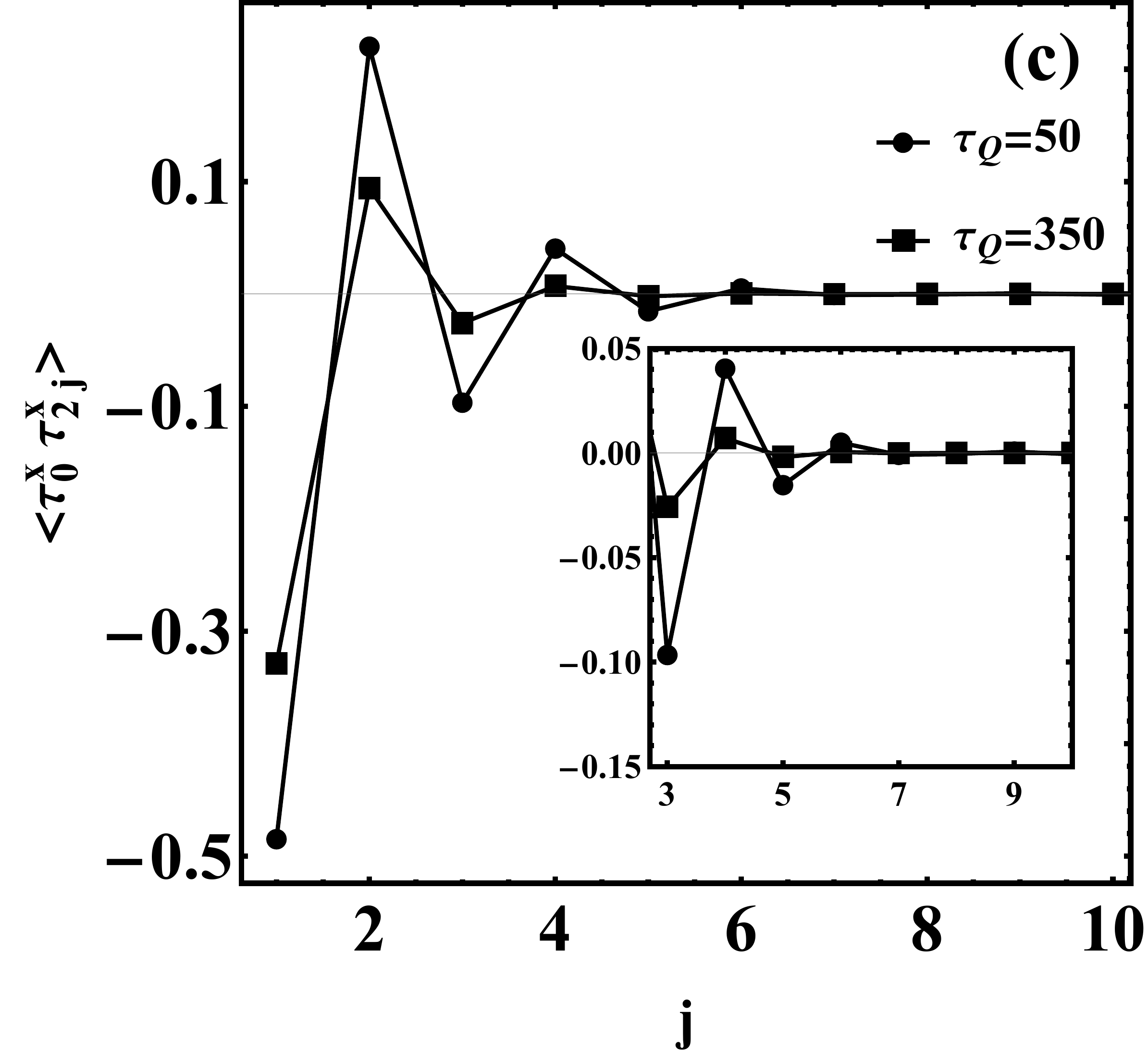}
     \caption{
     	 (a) $\langle \tau^x_0 \tau^x_{2j} \rangle$  vs. $j$ for the noisless scenario exhibits monotonic decay behavior at fast sweep speeds, at slower sweeps it exhibits oscillatory decay. 
(b)For fixed $\tau_Q=500$, the increased noise strength results in stronger decay. (c) For fixed noise strength $\eta_0 = 0.02$, in agreement with the AKZ behavior slower sweep speeds results in faster decay of the correlation.}
    \label{fig:hidden_corr}
\end{figure}

Performing the spin duality transformation~\cite{PhysRevD.17.2637, PhysRevLett.98.087204},  $\sigma^x_j =\tau^x_{j-1} \tau^x_j$ and $\sigma^y_j=\Pi^{N}_{l=j} \tau^{y}_l$,  the Hamiltonian as given in  Eq.~[\ref{eq:H_spin}] acquires the following form in the dual lattice space,
\begin{equation}
H_d= \sum^{N/2}_{j=1} (J_1 \tau^{x}_{2j-1} \tau^x_{2j+1}+J_2 \tau^y_{2j-1}), 
\label{eq:Hdual}
\end{equation}  
which is the usual one dimensional TFQIM. 

For $J_1 >J_2$ (i.e., $J_{-}>0$), the order parameter $\langle \tau^x_j \rangle$ acquires a finite value with respect to the ground state. Thus at the same time the expectation value of the operator $ \tau^x_1  \tau^x_{2j+1}$ with respect to the final decohered state acquires long range order implying the presence of hidden string order parameter in the original spin space.
 Our objective is to calculate the expectation value of this string operator
 \begin{equation}
\tau^x_1 \tau^x_{2j+1} = \Pi^{2j}_{l=1} \, \sigma^{x}_l=(-i)^j (a_1 b_1 a_2 b_2.....a_{j} b_{j}),
\end{equation}
where $a_j$ and $b_j$ are the Majorana fermions. As discussed above, the two point correlators which survive at the end of the protocol are of the form $\langle a_n b_{n+r}\rangle$. Therefore 
the  expectation value of the operator $\langle \tau^x_1 \tau^x_{2j+1}\rangle$  can be recast in to the Toeplitz matrix determinant form~\cite{LIEB1961407,PhysRevA.3.786},
\begin{equation}
\langle \tau^x_1 \tau^x_{2j+1}\rangle=(-1)^{j}\begin{vmatrix}
f_0 & f_{-1} &....& f_{-(j-2)} & f_{-(j-1)}\\
f_1 & f_{0} &....& f_{-(j-3)} & f_{-(j-2)}\\
. & f_1 &....& .&.\\
. & . &....& .&.\\
. & . &....& .&f_{-1}\\
f_{j-1} & f_{j-2} &....& f_{1} & f_{0}
\end{vmatrix},
\label{eq:T_determinant}
\end{equation}
where the elements of the Toeplitz matrix are given by,
\begin{equation}
f_r=\langle i a_n b_{n+r} \rangle\,\, \text{and} \,\, f_{-r}=\langle i a_{n+r} b_n \rangle. 
\end{equation}
 The numerically evaluated expectation values with respect to the finite separation have been plotted in the Fig.~[\ref{fig:hidden_corr}].
 One  notices that the expectation value of  $\langle \tau^x_0 \tau^x_{2j}\rangle$ shows monotonic decay at the fast sweep speeds and exhibits damped oscillatory behavior in the slower sweep regime.
  Meanwhile, in the noisy drive scenario, we find that the correlator is suppressed in the slow sweep regime with the increased noise strengths, as shown in the Fig.~[\ref{fig:hidden_corr}], which is in qualitative agreement with the AKZ picture.

\section{Kinks statistics}\label{KStat}
The kink number distribution $P(n)$ describes the probability of observing $n$  kinks. As discussed in the  previous section, one obtains TFQIM  after the spin duality transformation of the Kitaev model  as given by Eq.~[\ref{eq:Hdual}]. In the limit of large time $J_1$ remains greater than  $J_2$, therefore the system will be in the ferromagnetic state where the defects or kinks are the domain walls  in the dual spin space and the operator measuring the number of kinks can be written as,
\begin{equation}
\mathcal{N}=\frac{1}{2} \sum^{N/2}_{j=1}( 1-\tau^{x}_{2j-1} \tau^x_{2j+1}).
\label{eq:kink_operator}
\end{equation}
By considering the periodic boundary condition, $\tau^{x}_{N+1}=\tau^{x}_{1}$ (with assumption, i.e., $N/4$ is even) and applying the standard procedure of the JW transformation along with the Fourier transformation and the Bogoliubov transformation, the TFQIM  can be diagonalized yielding, $   H=\sum_{k}E_{k}( \gamma_{k}^\dagger\gamma_{k}+\gamma_{-k}^\dagger\gamma_{-k}-1)$,
where $\gamma_{k}$ are  the Bogoliubov modes satisfying the standard fermionic anti-commutation relation~\cite{dutta_aeppli_chakrabarti_divakaran_rosenbaum_sen_2015}. The energy for the $k$-mode is given by, $E_{k}= 2 \sqrt{J^2_{-}\sin^2{\frac{k}{2}}+J^2_{+}\cos^2{\frac{k}{2}}}$.
 Furthermore, the kink number operator in terms of the Bogoliubov modes has a simple representation as a sum of number of quasiparticles, $\mathcal{N}=\sum_{k>0} \gamma^{\dagger}_k \gamma_k$, where the average $\langle \gamma^{\dagger}_k \gamma_k \rangle$ is  the excitation probability of the quasiparticles,i.e.,  $p_k=\langle \gamma^{\dagger}_k \gamma_k \rangle$ with respect to the final state and will remain the  same as given by Eq.~[\Ref{eq:pk}] solved for the Hamiltonian Eq.~[\Ref{eq:H_matrix_noisy}] driven  linearly with respect to time by varying the parameter $J_{-}$ and noise in the $J_{+}$ parameter. 

 Following Ref.~\cite{Campo_2018}, the kink number distribution can be written as the  expectation value of the operator $\delta(\mathcal{N}-n)$ with respect to the the final state $\hat{\rho}_D$ at the end of the quench protocol,
 $
 P(n)=\text{Tr}\,[\hat{\rho}_D\,\delta(\mathcal{N}-n) ],
$
where the characteristic function  of $P(n)$ is defined by taking the Fourier transform of  $P(n)$,
\begin{equation}
\tilde{P}(\theta)=\int^{\pi}_{-\pi} d\theta P(n) e^{i n \theta}.
\label{eq:Pn_fourier}
\end{equation}
Furthermore, the cumulants generating function are obtained by taking the logarithm of the characteristic function, i.e., $\text{log}\, \tilde{P}(\theta)$, where  
$
\tilde{P}(\theta)=\Pi_{k>0} \left[ 1+ (e^{i \theta}-1) p_k\right].$
 The cumulants $\kappa_q$ of the distribution $P(n)$ can be obtained via the expansion of the cumulant generating function,
$
\text{log}\, \tilde{P}(\theta)=\sum^{\infty}_{q=1}\, \frac{(i \theta)^{q}}{q!}\,\kappa_q.
$
The cumulants of the Poisson binomial distribution function are given by,
\begin{equation}
k_q= (-i)^q \frac{d^q}{d \theta^q}\text{log}\,\tilde{P}(\theta)|_{\theta=0}.
\end{equation}
The first cumulant $\kappa_1=\langle n\rangle=\sum_{k>0} p_k$ is the mean, the second cumulant is the variance $$\kappa_2=\langle n^2\rangle-\langle n\rangle^2= \sum_{k>0} p_k(1-p_k),$$ 
 
 \begin{figure}[H]
 	\centering
 	\begin{subfigure}[!ht]{0.39\textwidth}
 		\centering
 		\captionsetup{justification=centering}
 		\includegraphics[width=\textwidth]{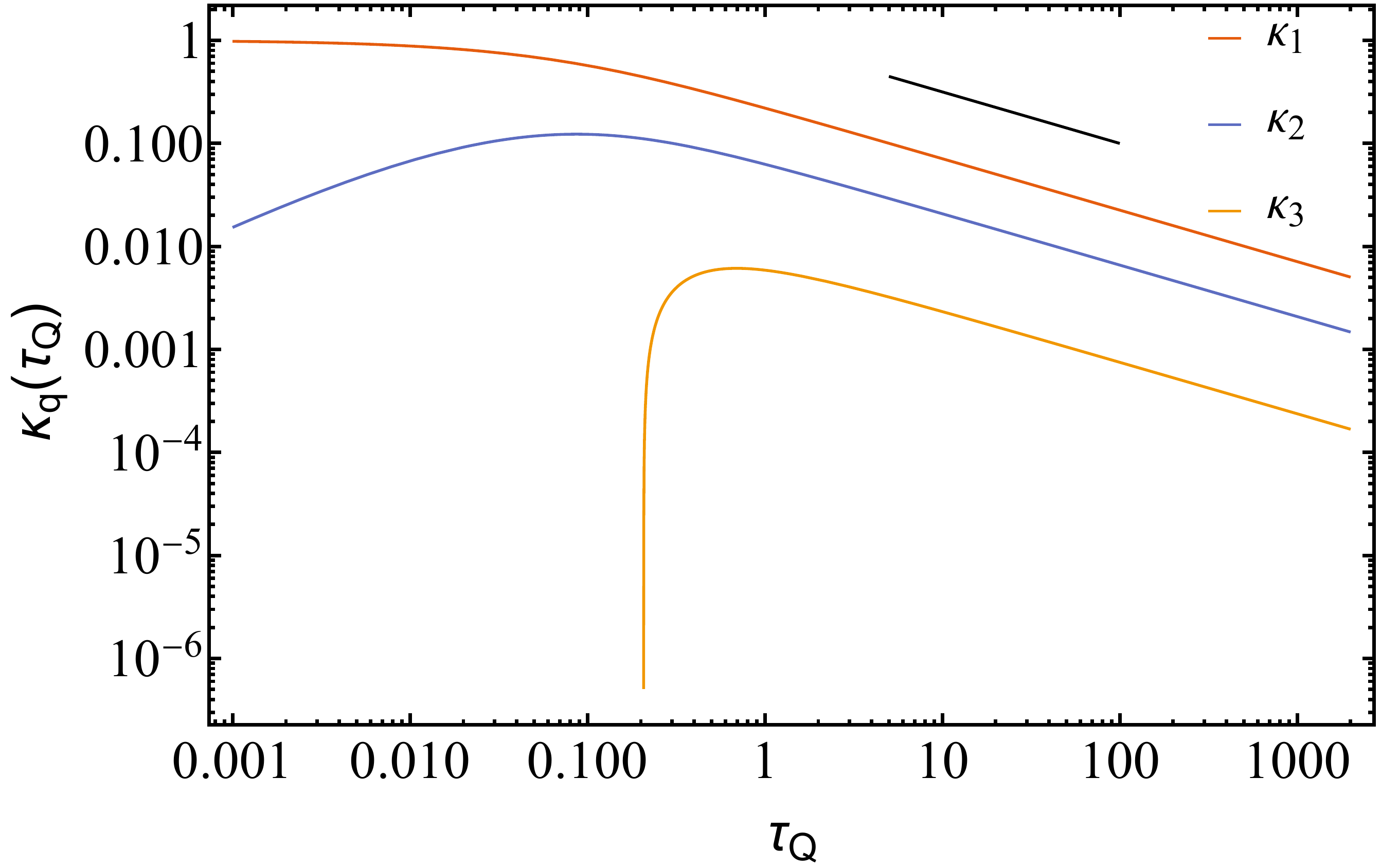}
 		\caption{For $\eta_0=0$ }
 		\label{fig:cumulants_analytical_1}
 	\end{subfigure}
 	\hfill
 	\begin{subfigure}[!ht]{0.39\textwidth}
 		\centering
 		\captionsetup{justification=centering}
 		\includegraphics[width=\textwidth]{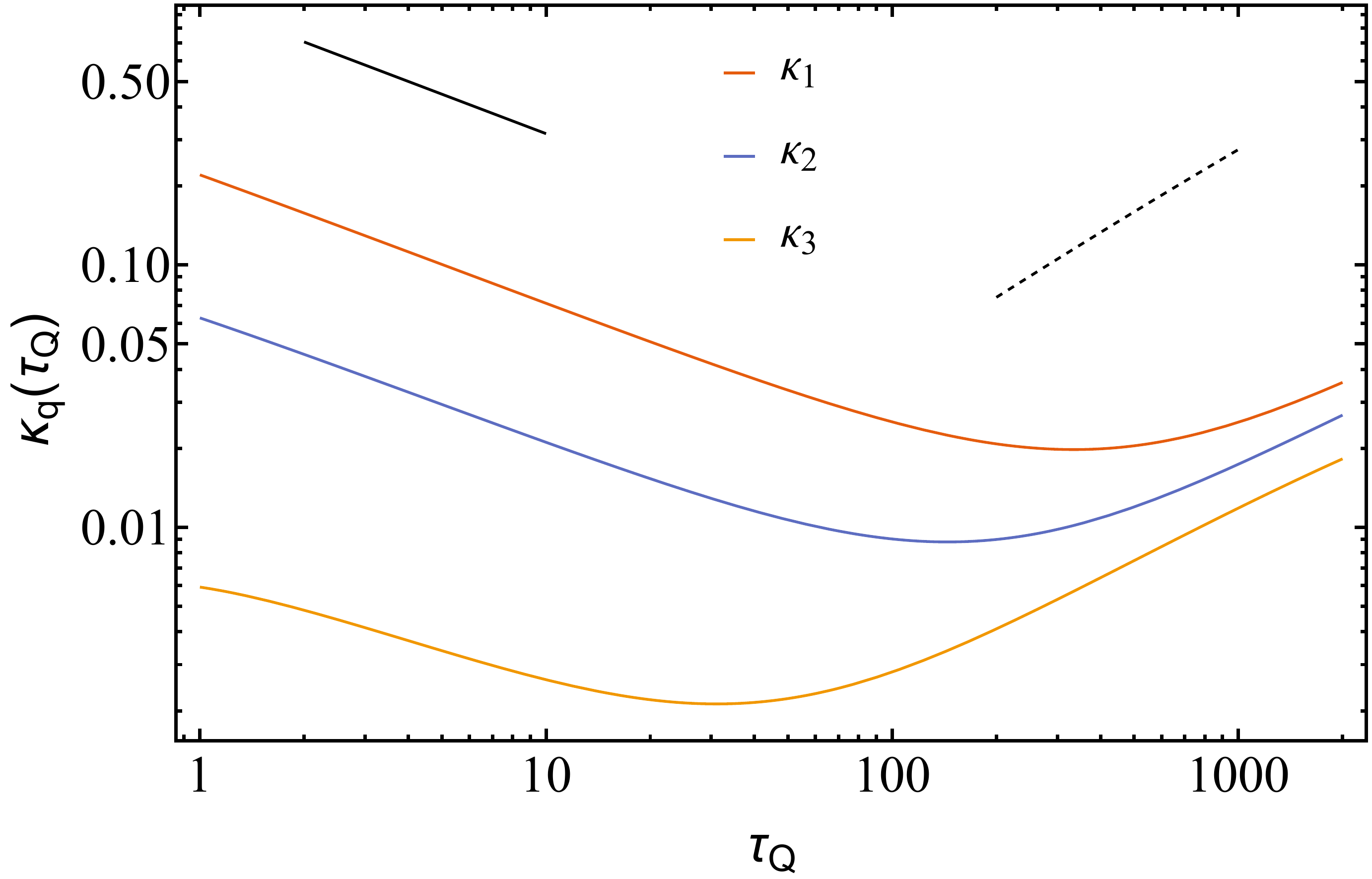}
 		\caption{For $\eta_0=0.001$}
 		\label{fig:cumulants_analytical_2}
 	\end{subfigure}
 	\centering
 	\begin{subfigure}[!ht]{0.39\textwidth}
 		\centering
 		\captionsetup{justification=centering}
 		\includegraphics[width=\textwidth]{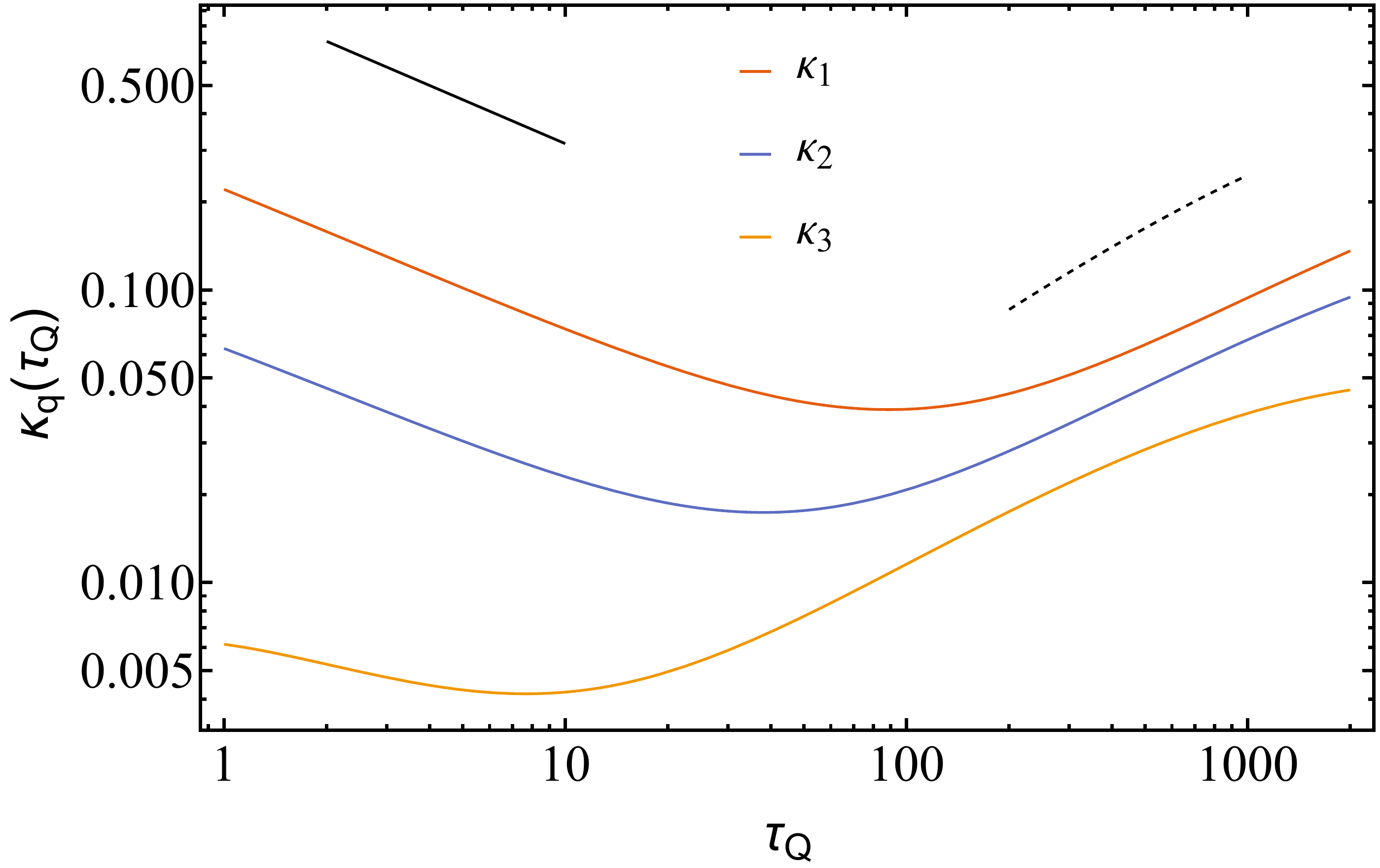}
 		\caption{For $\eta_0=0.003$}
 		\label{fig:cumulants_analytical_3}
 	\end{subfigure}
 	\hfill
 	\begin{subfigure}[!ht]{0.39\textwidth}
 		\centering
 		\captionsetup{justification=centering}
 		\includegraphics[width=\textwidth]{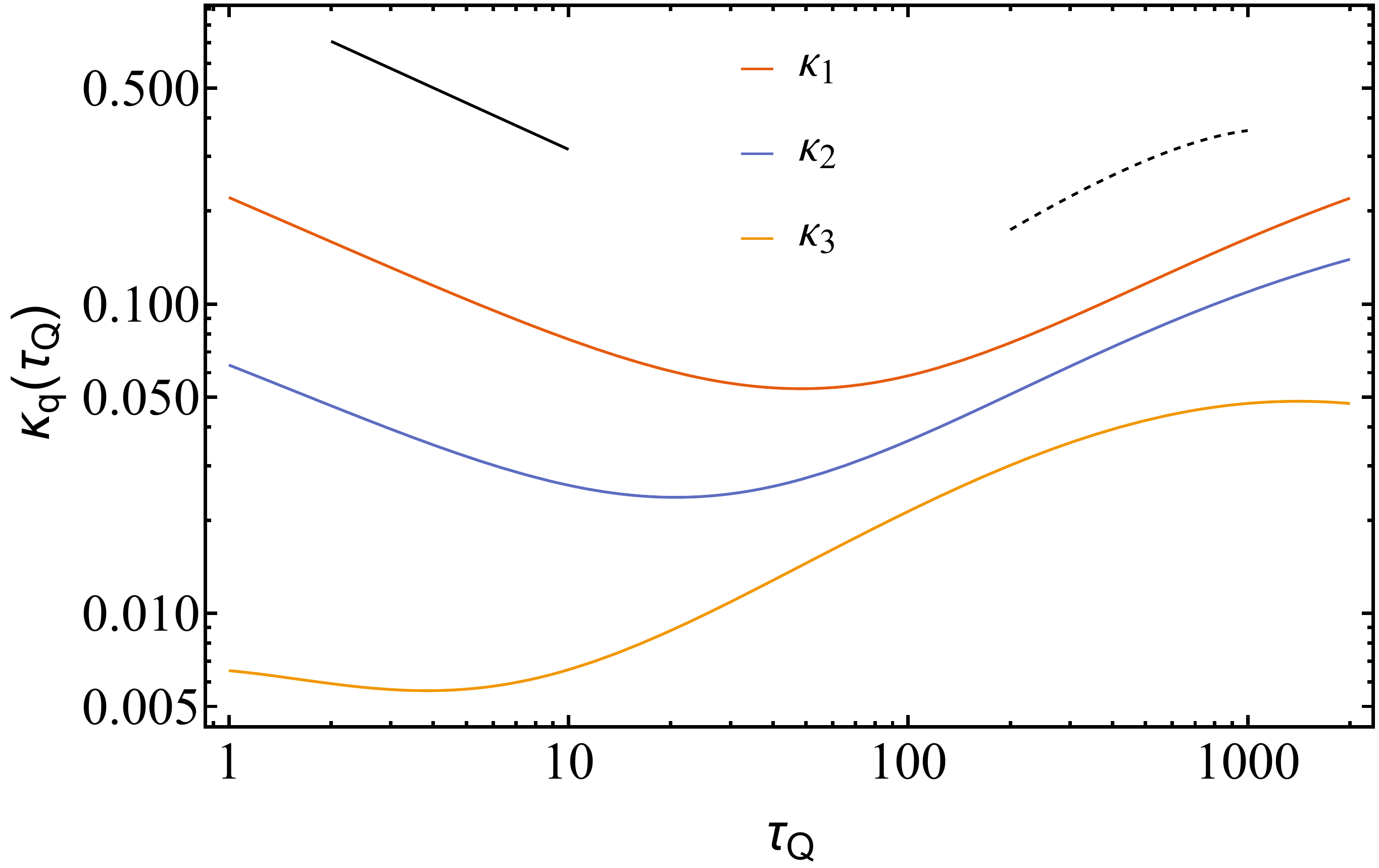}
 		\caption{For $\eta_0=0.005$}
 		\label{fig:cumulants_analytical_4}
 	\end{subfigure}
 	\captionsetup{justification=centering}
 	\caption{ Dependence of first three cumulants (density) of the kink number distribution on the quench time, $\tau_Q$, have been plotted for different noise strengths.  Here, the connected black line shows the slope proportional to $ 1/\sqrt{\tau_Q}$ while the dashed black line is proportional to $-\chi \log(\chi)$. As can be observed from (a) all three kink cumulants follow the universal KZ scaling behavior in the slow sweep regime as expected for the noiseless drive. The noisy drive scenarios as shown in (b), (c), and (d), the kink cumulants exhibit nonmonotonic behavior with the quench time consistent with the AKZ picture.} 
 	\label{fig:cumulants_analytical}
 \end{figure}
 and, $$\kappa_3=\langle (n-\langle n\rangle)^3 \rangle=\sum_{k>0} p_k(1-p_k)(1-2 p_k),$$ is the third central moment where $p_k$ is given by Eq.~[\ref{eq:pk}].  For the Gaussian distribution  the first two cumulants are non-zero while the higher cumulants vanish, i.e., $\kappa_q =0$ for $q>2$. 
Note that any non-zero higher cumulants tells us about the non-normal features of the  distribution. 
 In the thermodynamic limit,  by numerically  integrating over $k$, we plot  in Fig~[\ref{fig:cumulants_analytical}] the cumulants as a function of the quench time $\tau_Q$ for different noise strengths. For the noiseless limit the scaling behavior of the cumulants is consistent with the KZM picture, i.e., as shown in the Fig.~[\ref{fig:cumulants_analytical}](a) the first three cumulants exhibits $1/\sqrt{t_Q}$ universal KZ scaling with the quench time in the slow sweep regime as has also been observed in Refs.~\cite{King2022,Campo_2018,Cui2020}. However, the deviations from the KZ scaling behavior are apparent in the fast sweep regime when $\tau_Q <1$ (see Fig.~[\ref{fig:cumulants_analytical}](a)). 
Another reason  for the deviations from the KZM for the noiseless case can be attributed to a scenario  wherein the system size becomes comparable to the effective separation between two nearest defects~\cite{Campo_2018}. While the above argument is not relevant for our work, it acquires significance for experimental scenarios.  In the noisy drive, the cumulants exhibit  noise dependent AKZ scaling behavior proportional to the $-\chi \text{log}(\chi) $ term for slower sweeps (see  Fig~[\ref{fig:cumulants_analytical}]). We find analytically that the source of the anamolous term is the same as that for the defect density, i.e.,  from $\int dk p_k$.
 On the other hand,  the AKZ terms arising from  $\int dk p_k^2$ and $\int dk p_k^3$ are of the form,  $\chi$, which are   of the conventional type. For relatively fast sweep sweep speeds (quench time smaller than the optimal quench time, but still $\tau_Q >1$), the kink cumulants are unaffected by the noise and follows the KZ scaling behavior (similar to the case of coherent dynamics in an  the isolated quantum system~\cite{King2022,Cui2020}).
 
  To better understand the dependence of the kink cumulants on the quench time, we plot the ratios of the cumulants for different noise strengths in Fig.~[\ref{fig:cumulants_ratios}]. As can be observed in Fig.~[\ref{fig:cumulants_ratios}](a) for the noiseless case when $\eta_0=0$,  the cumulants  $\kappa_2 \propto \kappa_1$ and $\kappa_3 \propto \kappa_1$, specifically $\kappa_2/\kappa_1 =0.29$ and  $\kappa_3/\kappa_1=0.033$,  which turn out to be independent of the quench time in the slow sweep regime ($\tau_Q \gg 1$) as expected from the universal KZM scenario, i.e., $\kappa_q \propto t^{-1/2}_Q$ for all cumulants~\cite{Campo_2018, Cui2020, Bando2020, King2022}. 
 For $\eta_0=0.001$ case as seen from Fig.~[\ref{fig:cumulants_ratios}](b),  the ratios of the cumulants develop non-universal dependence on the quench time for slower sweeps and moreover, $\kappa_3/\kappa_1$ is affected first at relatively small quench time due to the noise which also shows that the higher cumulants are more susceptible towards noise, however the amplitude of the higher cumulants remain very small compared to the first two cumulants. Consider next  $\eta_0 =0.01$ (see  Fig.~[\ref{fig:cumulants_ratios}](c)) for which the ratio $\kappa_2/\kappa_1$ at small quench times are similar to that for the noisless case but increases at intermediate quench times finally saturating at $0.5$ for very long quench times. At the same time the ratio $\kappa_3/\kappa_1$   acquires a maxima (with quite small value as compared to the ratios $\kappa_2/\kappa_1$) at an intermediate quench time beyond which it monotonically  decreases with the ratio approaching zero. In the limit of large quench times and in the presence of noise the state of the system get maximally scrambled or $p_k \rightarrow 1/2$ leading to the above asymptotic values for the ratios of the cumulants.

 \begin{figure}
    \centering
    \includegraphics[width=0.49\columnwidth]{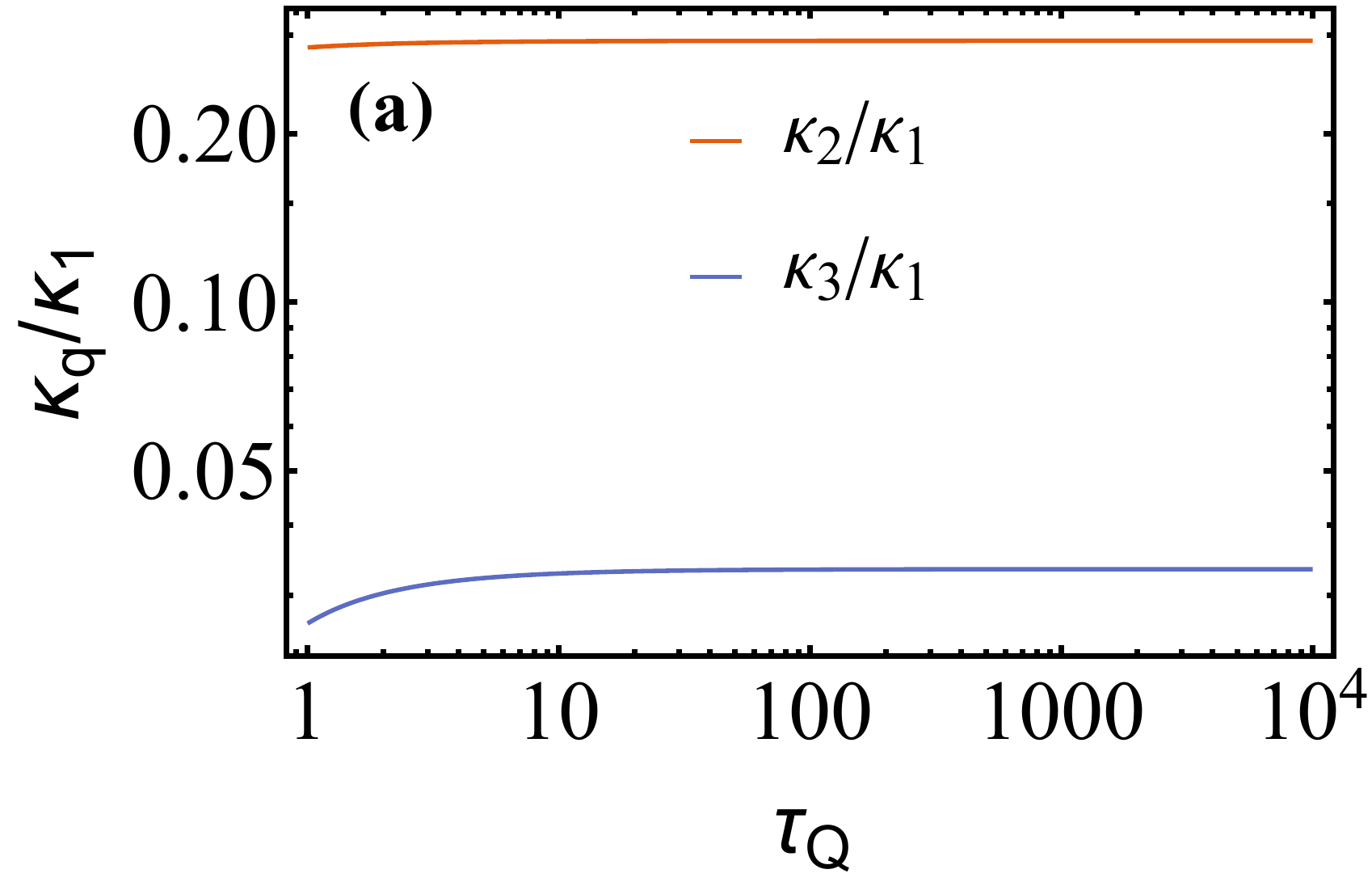}
    \includegraphics[width=0.49\columnwidth]{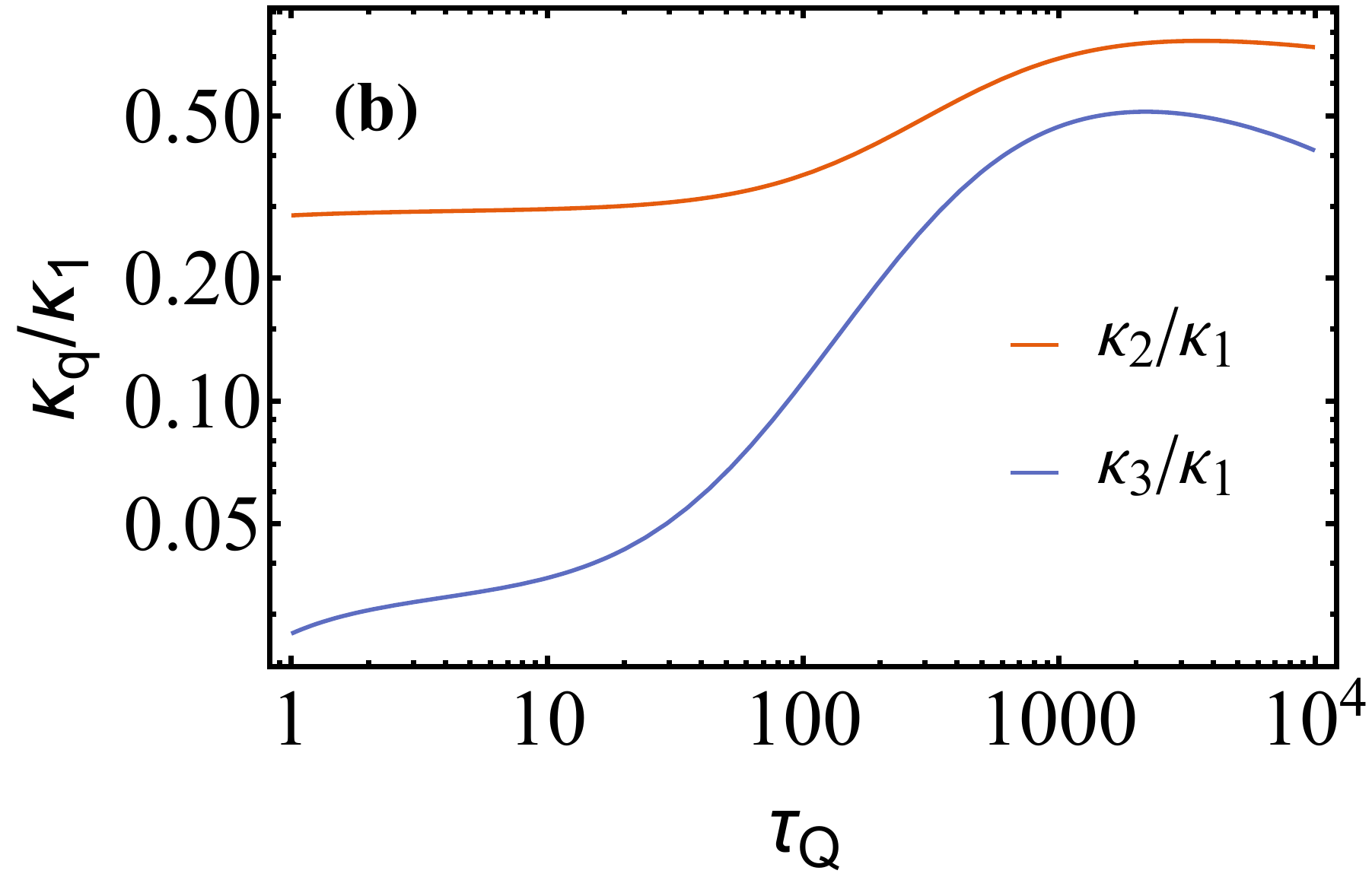}
    \includegraphics[width=0.49\columnwidth]{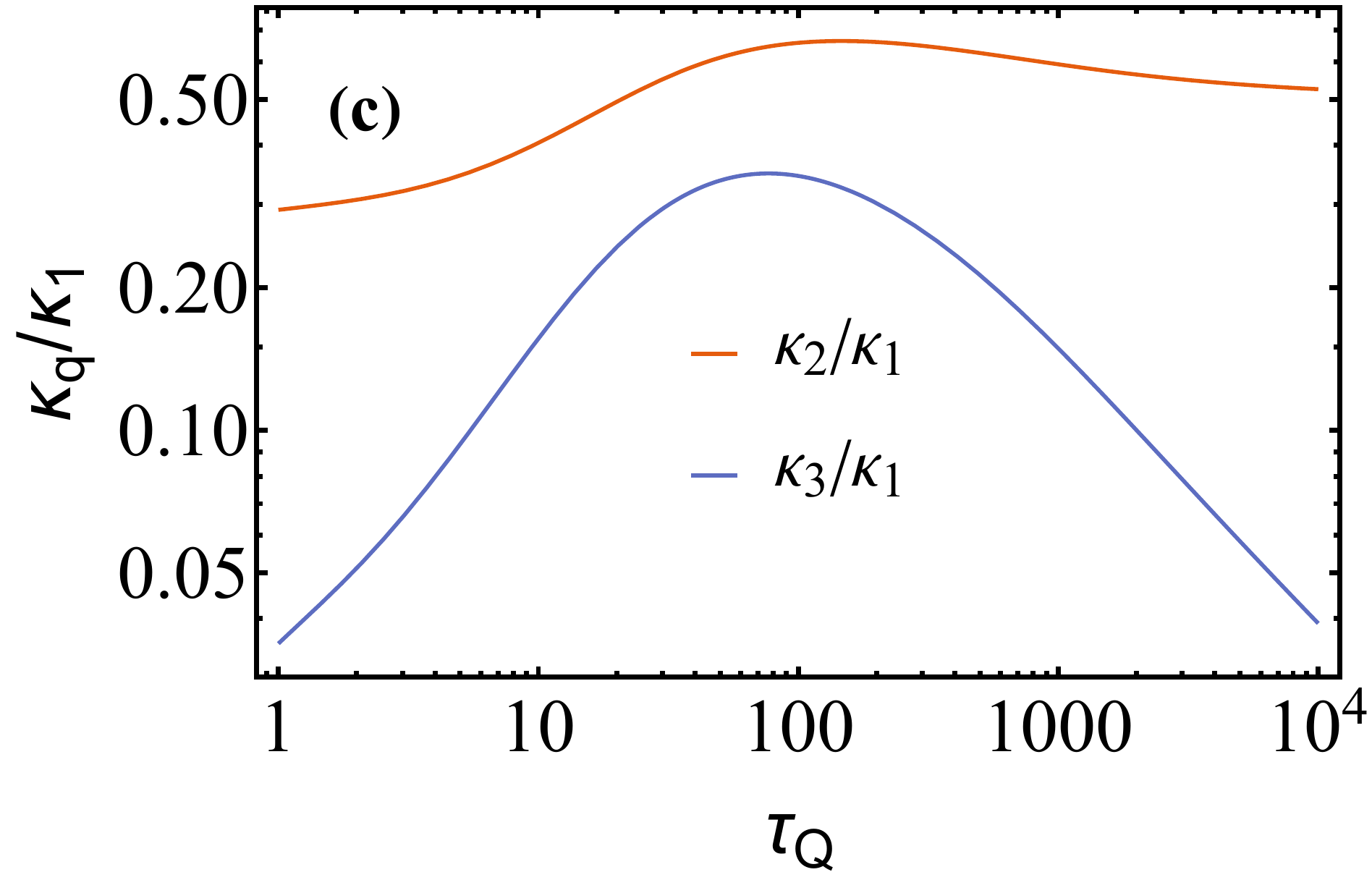}
     \caption{The ratios of the cumulants $\kappa_2/\kappa_1$ and $\kappa_3/\kappa_1$ have been plotted with the quench time for different noise strengths, i.e., (a) for $\eta_0=0$, (b) for $\eta_0=0.001$, and (c) for $\eta_0=0.01$ cases.}
 \label{fig:cumulants_ratios}
\end{figure}
\begin{figure}
    \centering
    \includegraphics[width=\columnwidth]{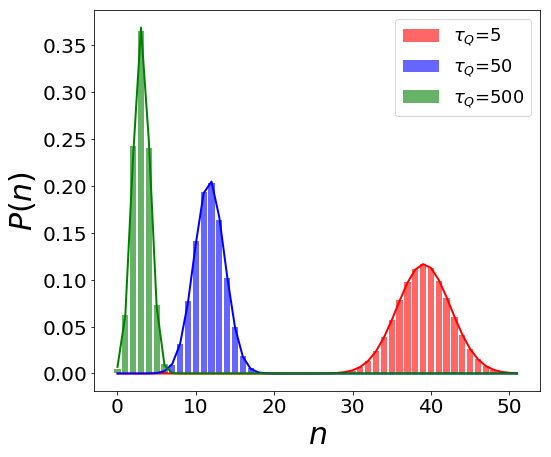}
     \caption{Plotted are the kink number distribution for different quench times (taking the system size- $N=800$).  The solid lines represents the Gaussian approximation, i.e., the normal distribution with mean $\kappa_1$ and variance $\kappa_2$.} 
    \label{fig:Kinks_distribution_noiseless}
\end{figure}
\begin{figure}
    \centering
    \includegraphics[width=\columnwidth]{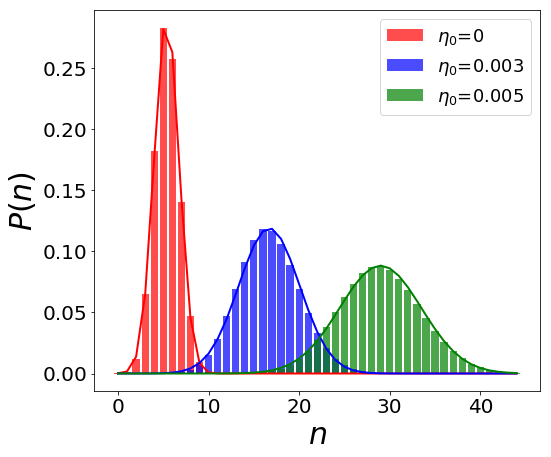}
     \caption{Kink distribution has been plotted for different noise strengths for quench time $t_Q=200$ (slow sweep regime). The solid lines represents normal distribution approximation with non-universal mean density $\kappa_1$ and variance $\kappa_2$, which depend on the quench time as well as the noise strength as described by the AKZ picture (see Fig.~[\ref{fig:cumulants_analytical}]). }
    \label{fig:Kinks_distribution_noisy}
\end{figure}
  Besides the cumulants,  we  numerically calculate the kink distribution  function by evaluating the  inverse Fourier transform of the characteristic function (Eq.~[\ref{eq:Pn_fourier}]). They are plotted in Figs.~[\ref{fig:Kinks_distribution_noiseless}] and [\ref{fig:Kinks_distribution_noisy}] for the noiseless drive scenario and for the noisy drive case, respectively. The kink distribution function is well approximated by the normal distribution $\mathcal{N}(\kappa_1,\kappa_2)$  with $\kappa_2 \propto \kappa_1$ (in spite of having nonzero higher cumulants $\kappa_q>0$ for $q>2$) for the noiseless case (see Fig.~[\ref{fig:Kinks_distribution_noiseless}]) and  away from the onset of the adiabaticity, i.e., $\langle n \rangle >1 $  and in slow sweep regime $\tau_Q >1$. This was first reported by the authors in Ref.~\cite{Campo_2018} and later verified in various experimental studies in Refs.~\cite{Cui2020, Bando2020, King2022}. For the noisy case as can  be seen from Fig.~[\ref{fig:cumulants_analytical}], the noise affects the cumulants most in the slow sweep
regime ($\tau_Q \gg 1$). We plot the kink distribution for different noise strengths at fixed quench time $\tau_Q=200$ (see Fig.~[\ref{fig:Kinks_distribution_noisy}]), the kink distribution of the noiseless limit shifts towards higher mean density values and broadens with increased variance values due to increased noise strengths consistent with the AKZ picture.  Interestingly, the kink distribution is still well approximated by the normal distribution (depicted by the solid lines) but with non-universal noise dependent mean $\kappa_1$ and variance $\kappa_2$ inspite of the noise dependent nonzero higher cumulants (since they still remain relatively quite small in comparison to first two cumulants even in the presence of noise for slow sweeps). However, note that $\kappa_2$ and $\kappa_3$ are no longer proportional to $\kappa_1$ unlike for the noiseless case.

The recent experimental studies simulating one dimensional TFQIM using superconducting flux quanta~\cite{King2022} and the quantum annealing protocol for the two dimensional TFQIM carried out with D-Wave device~\cite{Weinberg2020} have also found deviations from the universal KZM theory. The signature of the AKZ behavior are clearly visible in the kink density ($\kappa_1$) in the slow sweep regime (large annealing times) which can be attributed to nonzero temperature effects or due to the coupling of the system to noise source or environment. These studies indicate that the higher cumulants will also be susceptible to environmental noise or temperature specifically in the slow sweep regime,  which we have shown explicitly in Fig~[\ref{fig:cumulants_analytical}] considering the noisy drive through the QCP. In the  experimental work of  Bando et al.~\cite{Bando2020} the quantum annealing of the one dimensional TFQIM using the D-Wave device was studied and it was reported that the cumulants of the kink distribution follow  universal KZ scaling behavior  (with  $\kappa_q \propto  t^{-\alpha}_a$ ,  however the KZM exponent  $\alpha$ turns out to be sensitive to the noise-strength or environmental coupling).
 Interestingly, the absence of the nonmonotonicity in the kink density in their measurements was attributed to the fact that their annealing time range was too short for the noise  or temperature related effects to become significant at the end the drive protocol. All of the above experimental studies suggest that the effect of noise on the driven system strongly depends on the problem type as well as on the quench time range which is also verified from our present study.

\section{Discussion}\label{Disc}
In this article, we have studied driven one dimensional Kitaev chain in the presence of noise. In particular, the Hamiltonian is driven linearly with time by the parameter,  $J_{-}(t)$,  while the time dependent Gaussian noise is present in the parameter,  $J_{+}(t)$. In the quenching scheme involving the parameter  $J_{-}$, the quantum phase transition of topological type occurs at the QCP, $J_{-}=0$, which is related to the change of topological order across the QCP~\cite{PhysRevLett.98.087204, KITAEV20062}. The vanishing energy gap at the QCP leads to the diverging relaxation time which therefore results in the creation of defects/excitations across the QCP at the end of the quench protocol. We have analytically calculated the momentum-dependent LZ excitation probability for the quenching scheme by $J_{-}$ at the final time ($t\rightarrow \infty$) in the limit of fast noise which clearly differs from the transverse drive protocol of the TFQIM or the XY chain, i.e., here the new non-adiabatic region opens up around $k=0$ instead of the  $k=\pi/2$ region.
This has been utilized to evaluate the defect density numerically and also analytically which shows the AKZ  behavior, i.e., the slower noisy drives creating more defects however the AKZ term, $-\chi \text{log} \chi $, in weak noise limit  differs from the usual AKZ scaling behavior reported in all the previous studies. 
As a result, the optimal quench time required for the minimization of the defect density exhibits a rather complicated power law scaling behavior with the noise strength given by Eq.~[\ref{eq:n_optimal}] (see also Fig.~[\ref{fig:n}](b)). The residual energy calculated in the final decohered state also exhibits the AKZ  behavior with AKZ term proportional to $\chi$ resulting in the conventional power law scaling with the noise strength of the optimal quench time scale associated with the minimized energy density, i.e., $\tau^{e}_O \propto \eta^{-4/3}_0$ (see Fig.~[\ref{fig:n}](d)).
In addition, the numerically evaluated  von Neumann entropy density which gives the measure of decoherence due to the drive through the QCP exhibits 
a local maxima  at an  intermediate quench time scale, either side of which the entropy density decreases.   In the large quench time regime, the von Neumann entropy exhibits an upturn consistent with the AKZ behavior arising due to the noise. This behavior of the entropy density  can be linked with the crossover behavior of two-point Majorana correlator at zero separation.

It turns out that certain non-local string correlators can capture the hidden topological order across the QCP. 
The two-point Majorana correlator and  the hidden string correlator exhibit signature of  KZ scaling behavior in the noiseless case i.e.,  slower drive leads to larger correlation lengths. In particular, we also analytically approximate the two-point Majorana correlator in the limiting cases  for $\chi \ll 1$ and $\chi \gg 1$ as given by Eq.~[\ref{Eq:O_rweak}] and Eq.~[\ref{Eq:O_rstrong}], respectively.
We next consider the expectation value of the hidden string correlator with respect to the final decohered state which can be written as product of two-point Majorana correlators.  The product in turn can be expressed in the Toeplitz matrix determinant form. Evaluating the Toeplitz determinant numerically, we find that the hidden string correlator exhibits signature of the AKZ picture of defect generation i.e. the correlator becomes short ranged in the presence of noise for the slow drive case. 
   The analysis of kink statistics is performed in the dual space, where cumulants till the third order were considered.   
  	These also exhibit AKZ scaling behavior in the noisy drive scenario similar to the defect density in  the weak noise limit, i.e., dominant contribution comes from the AKZ term proportional to, $-\chi \text{log} \chi $  in $\chi \ll 1$ limit for slower sweeps as shown in Fig~[\ref{fig:cumulants_analytical}]. Furthermore, the kink distribution is well approximated by the normal distribution with non-universal noise dependent mean density and variance for slower sweeps as shown in Fig.~[\ref{fig:Kinks_distribution_noisy}], but unlike the KZM scenario of noiseless drive case, for the noisy case the higher cumulants are not proportional to the first cumulant as can also be seen from Fig.~[\ref{fig:cumulants_ratios}].

\section{ Acknowledgement}
 S. G.  and M. S. are grateful to  the SERB, Government of India, for the support
via the Core Research Grant  Number CRG/2020/002731. We would like to thank  Prof. S. Chaturvedi and Prof. D. Sen for the useful discussion.

%

\end{document}